\documentclass[aps,physrev,twocolumn,superscriptaddress]{revtex4-2}
\usepackage{graphicx}% Include figure files
\usepackage{dcolumn}% Align table columns on decimal point
\usepackage{bm}% bold math
\usepackage{amsmath}
\usepackage{circuitikz}
\usepackage{overpic}
\usepackage{mathptmx}
\usepackage{etoolbox}
\usepackage{hyperref}
\hypersetup{colorlinks,allcolors=black}

\begin{document}

%\preprint{}

\title{Sub-kelvin thermal conductivity of substrates and on-chip routing in quantum integrated systems}


\author{Charles Bon-Mardion}
\email{charles.bonmardion@gmail.com}
%\thanks{}
\affiliation{Univ. Grenoble Alpes, CEA, LETI, Grenoble, France}
\author{Arnaud Lorin}
\author{Edouard Deschaseaux}
\author{Céline Feautrier}
\author{Daniel Mermin}
\author{Jean Charbonnier}
\author{Jing Li}
\affiliation{Univ. Grenoble Alpes, CEA, LETI, Grenoble, France}

\author{Jean-Luc Sauvageot}
\affiliation{Univ. Paris-Saclay, CEA, IRFU, Gif-sur-Yvette, France}
\author{Candice Thomas}
\email{candice.thomas@cea.fr}
\affiliation{Univ. Grenoble Alpes, CEA, LETI, Grenoble, France}


\date{May 7, 2026}

\begin{abstract}
%The development of quantum technologies increasingly depends on optimized thermal management at cryogenic temperatures. 
The development of large-scale quantum systems increasingly relies on the close integration of heterogeneous components such as qubits, control electronics, and readout circuits, making thermal management at cryogenic temperatures a central challenge in such architectures.
In this work, we present an experimental thermal study of two building blocks of such systems: the substrate and the on-chip routing.
We first investigate the sub-kelvin thermal conductivity of four substrate materials: high-resistivity silicon, low-resistivity silicon, borosilicate, and sapphire. 
We report that high-resistivity silicon exhibits the highest thermal conductivity among the substrates studied ($5\cdot10^{-2}$~W/m$\cdot$K at 300~mK), while low-resistivity silicon, borosilicate, and sapphire show lower values ($8\cdot10^{-4}$~W/m$\cdot$K, 2$\cdot10^{-3}$~W/m$\cdot$K, and 2$\cdot10^{-3}$~W/m$\cdot$K at 300~mK, respectively). 
Ballistic conductance evaluation using a finite-element non-equilibrium Green's function approach further allows us to extract the phonon mean free path in each substrate and gives insights into the involved scattering mechanisms. %identify the dominant scattering mechanisms.
Additionally, we employ a dedicated test vehicle to evaluate the impact of on-chip routing on the thermal conductance of the system. 
Our measurements with superconducting Nb routing lines reveal that the routing increases the in-plane thermal conductance of the system, but the substrate remains the dominant heat path. 
These results highlight the critical role of the substrate choice within quantum systems and underscore the importance of function partitioning through 3D integration approaches for more efficient thermal management in quantum architectures.
\end{abstract}

%\keywords{}

\maketitle

%------------------------------
%--------Section---------------
%------------------------------

\section{Introduction}
The transition to fault-tolerant quantum computers inherently requires a massive scaling of the number of physical qubits \cite{Q_advantage_requireement}.
Typically, classical control electronics are located outside the dilution refrigerator and connected to the quantum layer via extensive coaxial cabling. However, scaling to millions of qubits renders this approach unmanageable, driving the need to place classical cryogenic complementary metal-oxide-semiconductor (cryo-CMOS) control circuits in close proximity to the qubit layer \cite{Summary_perspective_2025}. 
%either by integrating them on the same substrates in a system-on-chip approach, or by fabricating them on different wafers with potentially different technology nodes and connecting them through 3D integration techniques such as the use of an interposer as sketched in Fig1. In the following, we will refer to the first approach as "system-on-chip" and the second one as "system-in-package".
These highly integrated architectures generally follow either a system-on-chip (SOC) approach, where qubits and control electronics circuits are fabricated on the same wafers with the same technology \cite{REVIEW_3D_Integration_for_quantum,Monolithic_2024,Quobly_monolithic_2025}, or a 3D system-in-package (SIP) implementation, which relies on fabricating them on different wafers, with potentially different technology nodes and material platforms, and connecting them through three-dimensional (3D) integration techniques \cite{REVIEW_3D_Integration_for_quantum,Candice_quantum}.
Alternatives approaches also exist, where the control electronics are placed at a higher temperature stage of the dilution refrigerator (typically the 4~K stage) \cite{2026_Si_Quantum_HRL,2025_Rev_Classical_interfaces}, benefiting from the larger cooling power available compared to the sub-kelvin stage \cite{2026RessourcesEstimationFTQC}. However, this comes at the cost of increased wiring complexity and signal latency between stages. 
In all these paradigms, thermal management emerges as a major physical constraint \cite{Reilly2019,2024Scientific_challenges_cryoCMOS,2026RessourcesEstimationFTQC}. The heat dissipated by the cryo-CMOS (typically several microwatts per addressed qubit \cite{Cryo-CMOS_WATT}) introduces decoherence sources that degrade qubit lifetimes.
Understanding the thermal properties of the fundamental building blocks of quantum systems is thus a critical prerequisite for the design of large-scale quantum processors with optimized thermal management.

At cryogenic temperatures, thermal transport behaves differently than at room temperature. In dielectrics, heat is primarily carried by phonons and their propagation is mainly limited by impurities and lattice defects. This holds true until the temperature drops sufficiently low that the phonon mean free path approaches the physical dimensions of the sample itself (e.g. reaching centimeter scales below 2~K in silicon \cite{Si_MFP_lowT}).
In this boundary scattering regime \cite{Casimir}, thermal transport is no longer solely an intrinsic material property but becomes heavily dictated by sample size, geometry, and surface properties.
In parallel, while conduction electrons provide an additional heat channel in metallic routing layers, the integration of superconducting material suppresses this contribution. 
Thus thermal transport within quantum architecture with superconducting routing is predominantly governed by phonon propagation. 
While the thermal conductivity of common bulk materials is well-documented from room temperature down to a few kelvin \cite{ThermalCondOfEl1972}, systematic experimental data in the sub-kelvin regime remain scarce.
Because thermal transport at these temperatures is exquisitely sensitive to crystalline purity and sample dimension, high-temperature bulk values cannot be reliably extrapolated. 
Accurate thermal modeling of quantum architectures therefore requires empirical characterization of the materials involved, and in representative geometries. 
Given the complexity of such architectures, cryogenic thermal studies have focused so far on specific blocks, including: indium bumps 3D interconnects \cite{VTT_superconducting_assemblies}, superconducting circuits on silicon substrate \cite{Pekola_Thermal_Budget}, on-chip heating effects on a Si/SiO$_2$ chip \cite{Overheating_Si_2022}, and transistor self-heating \cite{TransistorHeating_1,TransistorHeating_2}. However, interpreting the thermal behavior of these complex integrated structures first requires an understanding of the main building blocks: the substrate and the routing lines.

This work addresses this need in two steps. First, we measure the sub-kelvin effective thermal conductivity of several substrate materials: silicon, borosilicate, and sapphire. Those substrates are used in standard microelectronics and also for quantum technologies \cite{Sapphire_Wang2022,Sapphire_23,Silicon_Bland2025,Boro_quantum_1,Boro_quantum_2}. From the obtained thermal conductivities, an analysis of the phonon scattering mechanisms is then provided through mean free path (MFP) extraction.
In the second part, we extend this study to a simple on-chip system comprising superconducting Nb routing layers on top of a Si substrate to quantify the impact of such integration on the in-plane thermal conductance. Ultimately, our findings provide quantitative guidance for substrate selection and underscore the critical necessity of thermal decoupling for future quantum and more globally cryogenic systems.

%----------------------------------------------------------------------------------------
%	SECTION 
%----------------------------------------------------------------------------------------
\section{Substrate thermal conductivity study}\label{Sec:Substrats}

Evaluating the thermal properties of the substrate is essential, as it typically represents the largest volumetric component of any integrated circuit. It is also the main thermal link between active components, those responsible for heat dissipation, and the cryostat or whatever lies in between. 
In this section, the thermal conductivity of four different substrates, commonly used in semiconductor process technologies, is measured at sub-kelvin temperatures, and the data are compared in light of our targeted application. The phonon mean free path is then extracted from these measurements to identify the dominant phonon scattering mechanism in each substrate.

\subsection{Substrate samples description}
The selection of a substrate for quantum architectures is driven by a trade-off between radio-frequency (RF) performance \cite{CEA_RF_substrats}, thermal management, and industrial manufacturability. 
Accordingly, the substrates investigated in this work comprise two silicon variants, borosilicate glass, and sapphire.
For superconducting qubits particularly, substrate selection is critical, as material losses directly impact coherence lifetimes.
Historically, sapphire has long been considered as the best platform to improve superconducting qubit lifetimes \cite{Sapphire_Wang2022,Sapphire_23}. Yet its incompatibility with large-scale CMOS fabrication lines remains a major bottleneck for scaling \cite{Sapphire_limitation}. A recent shift, however, has demonstrated record qubit lifetimes using high-resistivity silicon substrates \cite{Silicon_Bland2025}. This demonstration opens the way to build large scale qubit platforms using manufacturing process heavily optimized for silicon wafers.

In this work, two silicon variants are thermally characterized: boron-doped, low-resistivity (LR) silicon grown by the Czochralski (CZ) method, and high-resistivity (HR) silicon produced by the float-zone technique. 
While {CZ} wafers dominate in large-scale industrial applications, they are typically characterized by an elevated oxygen concentration due to the dissolution of silica from the crucible into the molten silicon during growth \cite{Si_CZ_Fz}. On the contrary, the float-zone method allows a crystal formation of the silicon that is kept untouched by any material during the growth, drastically increasing the purity of the crystal but at a higher cost \cite{Si_CZ_Fz}. 

In addition to the crystalline substrates, we also choose to investigate borosilicate glass as a representative amorphous alternative, available in 200~mm wafer format.
Although its application is highly specialized and not typically intended for solid state qubits, borosilicate glass remains relevant for integrated photonic circuits within quantum technologies \cite{Boro_quantum_1,Boro_quantum_2}. 
%Furthermore, it presents a interesting substrate option due to its optical transparency and excellent low-loss electrical properties \cite{RF_Glass_review}.

The essential physical properties of the studied substrates are synthesized in Table \ref{table:substrats}. Values were extracted from wafers manufacturer datasheets \cite{Si_Cz_Wafer, Si_FZ_Wafer, Borofloat_wafer, Sapphire_wafer}.
Both the silicon and the borosilicate samples were diced from their original 200~mm wafer into 34$\times$17~mm$^{2}$ pieces. The sapphire sample, obtained from a 50~mm wafer, differs in size and was prepared as a smaller $25\times12$~mm$^2$ piece.

\begingroup
\renewcommand{\arraystretch}{1.2} %Default value: 1 
\begin{table*}
    \caption[Physical properties of the substrate materials used in this study.]{Substrate properties. The silicon and borosilicate samples are taken from 200~mm wafers, the sapphire sample comes from a 50~mm wafer. Sample thicknesses were measured using an absolute length gauge. All other material parameters were taken from manufacturer datasheets: LR silicon \cite{Si_Cz_Wafer}, HR silicon \cite{Si_FZ_Wafer}, borosilicate glass \cite{Borofloat_wafer}, and sapphire \cite{Sapphire_wafer} wafers. }  %(Heidenhain AT1218)
    \centering
    \begin{tabular}{l cccc}
    \hline\hline
       Material &  LR Silicon  & HR Silicon  & Borosilicate  & Sapphire  \\
    \hline
        %Type & Single-crystal & Single-crystal & Amorphous & Single-crystal \\
    
       Crystal orientation & $<100>$ & $<100>$ & Amorphous & $<0001>^a$ \\
    
        Doping & Boron (p-type)  & Boron (p-type)  & - & - \\
        %p-type
        Resistivity ($\Omega\cdot$cm) & $1\text{--}50$ & $16 \cdot 10^3\text{--}19\cdot 10^3$ & $>10^{14}$ & $>10^{14}$ \\

        Oxygen (cm$^{-3}$) & $<7.3\cdot10^{17}$ & $<5\cdot10^{15}$  & -  & -  \\
    
        Carbon (cm$^{-3}$) & $<1.5\cdot10^{16}$ & $<1.5\cdot10^{16}$ & -  & -  \\
    
       Thickness $d$ (µm) & $731$ & $727$ & $730$ & $536$ \\
    
       Length $L$ $\times$ Width $w$ (mm) $^b$ & 34$\times$17 & 34$\times$17 & 34$\times$17 & 25$\times$12 \\ 
    \hline\hline
    %   Width (mm) & 17 & 17 & 17 & 12 \\
    %\hline
    \multicolumn{5}{l}{$^a${C plane (0001) to M (1-100) with a $0.2^{\circ}\pm0.1^{\circ}$ off}}\\
    \multicolumn{5}{l}{$^b${Length and width of the samples}}\\
    \end{tabular}
    \label{table:substrats}
\end{table*}
\endgroup

\subsection{Experimental setup and methods}
\begin{figure}
 % ---- First figure with (a) ----
    \centering
    \begin{overpic}[width=0.95\linewidth]{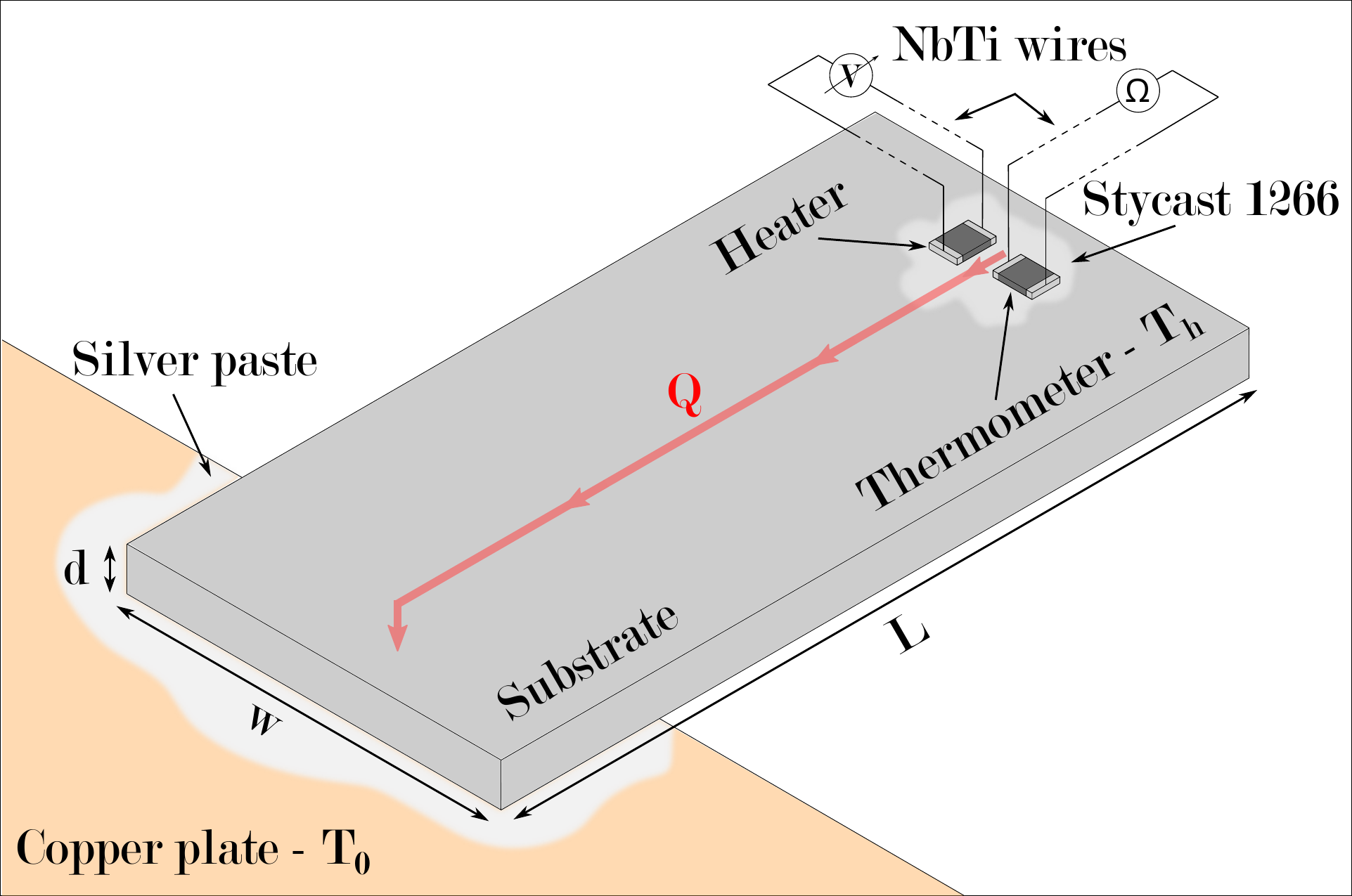}
        \put(3,61){(a)} % adjust coordinates if needed
    \end{overpic}
    % ---- Second figure with (b) ----
    \begin{circuitikz}[line width=1pt]
    \ctikzset{bipoles/thickness=1.2}
     % Label (b) top-left
    \node at (-8,1) {\small (b)}; % adjust coordinates as needed
    % Input node T0
    \draw (-8,0) node[circle, fill=black, inner sep=1.5pt] {} 
          node[above=0.1cm] {$T_{0}$};

    % Resistor chain with more spacing (reversed order)
    \draw (-8,0) to[R, l={$R_{b\,Substrate/Cu}$}] (-5,0)
          to[R, l={$R_{Substrate}$}] (-3,0)
          to[short, i<^={Heat $Q$}] (-1,0); % Arrow reversed
    
    % Vertical T_h line
    \draw (-3,0) -- (-3,0.6) node[circle, fill=black, inner sep=1.5pt] {}
          node[above=0.6cm] at (-3,0) {$T_{h}$};

    \end{circuitikz}
    \caption{(a) Experimental setup for the thermal conductivity study of various substrates. (b) Simplified equivalent thermal model of the setup.}
    \label{fig:Substrat_Sample}
\end{figure}
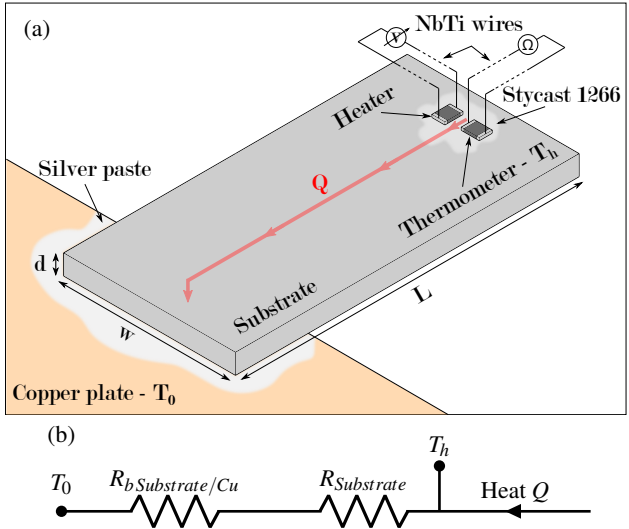

The measurement setup for the thermal conductivity characterization of the substrates is presented in Fig.~\ref{fig:Substrat_Sample}(a) where $L$ is the length, $w$ the width and $d$ the thickness of the sample. 
To ensure efficient thermalization of the cold-end side of the substrate, a substantial portion of the sample surface is glued on a copper plate using silver paste. This copper plate is screwed to the mixing chamber plate, ensuring proper thermal anchoring to the cryostat. For the silicon and borosilicate samples, the contact area is roughly $\sim$136~mm$^2$, while it is closer to $\sim$100~mm$^2$ for the sapphire sample.
In this configuration, the right-hand part of the sample is suspended, to force the thermal energy flows from the hot end at $T_h$ through the cross-section of the structure in the direction of the thermal bath at $T_0$.
For the measurement, a RuO$_2$ thermometer and a heater, both in a surface mounted device ({SMD}) packaging, are glued to the right edge of the sample using an electrically insulating epoxy (Stycast~1266). Physically the two components are in close contact, with a distance between them of the order of 1 to 2~mm and are not in a galvanic contact.
These components are connected to the measurement electronics using long superconducting NbTi wires ($\sim$20~cm), whose ends are thermalized at the cryostat base temperature $T_0$.
The heater is biased, and the thermometer resistance is measured using a cryogenic AC resistance bridge with a 300~µV excitation. Ensuring that the self-heating of the thermometer will be negligible relative to the lowest applied heating power (1~nW) used for the measurements.

The considered thermal equivalent model is presented in Fig.~\ref{fig:Substrat_Sample}(b) where $R_{Substrate}$ corresponds to the thermal resistance arising from the substrate sample material itself whereas $R_{b\,Substrate/Cu}$ designates the thermal boundary resistance between the sample and the copper plate. 
Note that, for simplicity, several components are omitted from the considered model.
First, the parallel thermal resistance of the NbTi wires is ignored, as their thermal resistance vastly exceeds that of the sample \cite{NbTi-CuNi_wires}.
Additionally, the intrinsic thermal resistance of the small Stycast volume, along with its associated thermal boundary resistance at the sample interface, is assumed to be negligible relative to $R_{Substrate}$.

To accurately extract the substrate's effective thermal conductivity from the total measured resistance $R = R_{Substrate} + R_{b\,Substrate/Cu}$, we must first ensure that the sample/copper boundary does not act as a thermal bottleneck (i.e, $R_{b\,Substrate/Cu} \gg R_{Substrate}$).  
The dominant thermal resistance must be the substrate itself (i.e, $R_{b\,Substrate/Cu} \ll R_{Substrate}$).
To verify this, a control experiment utilizing a second thermometer on the sample's cold side was performed, as detailed in the supplemental material \cite{SM}.
The test confirmed that the silicon/copper interface did not impose any thermal bottleneck on the overall measured thermal resistance. Because the copper/substrate thermal boundary resistance is not expected to vary significantly across the materials under study \cite{SwartzPohl}, the same conclusion was extended to all samples.

\begin{figure}
    \centering
    \includegraphics[width=0.95\linewidth]{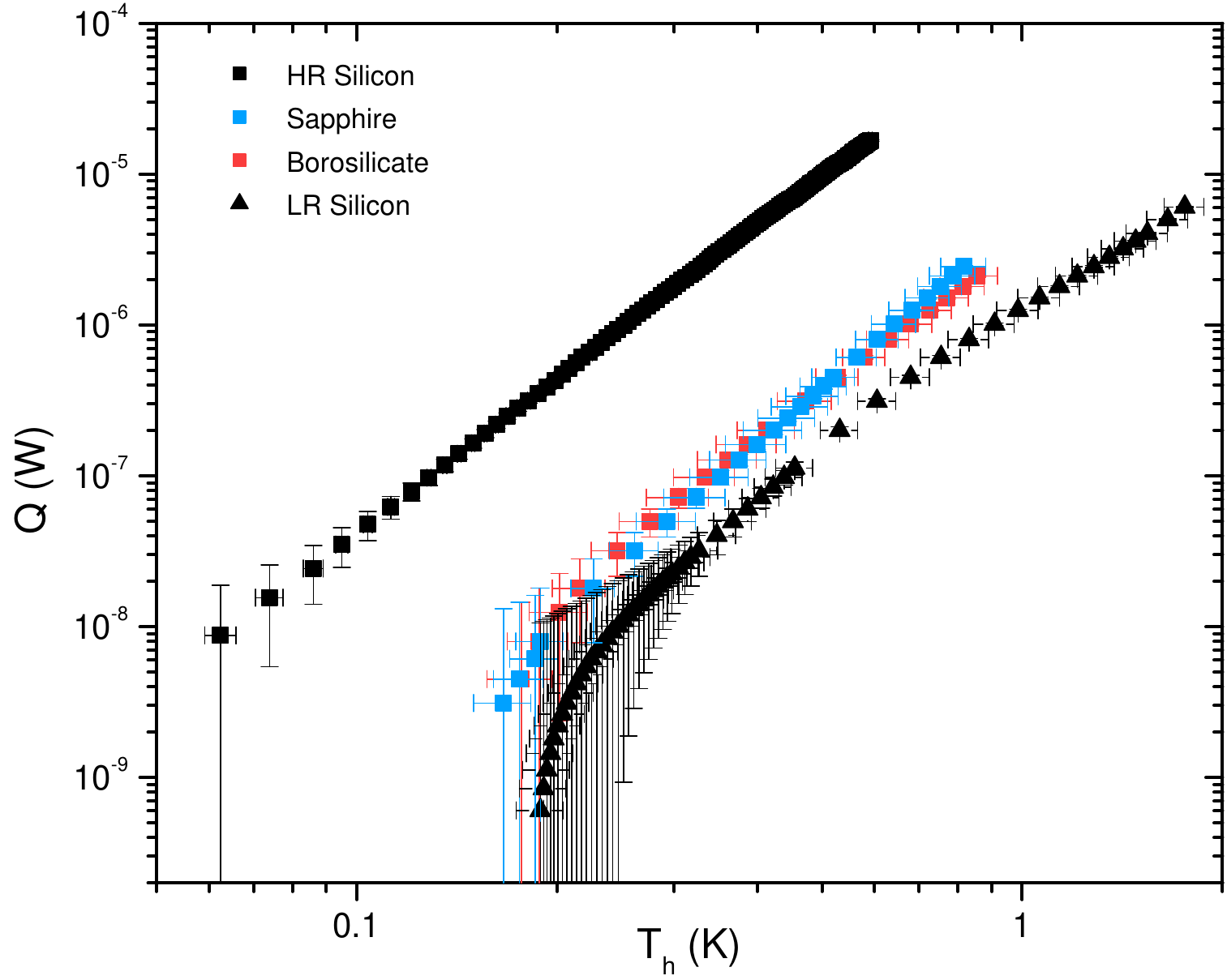}
    \caption{Applied power $Q$ versus the hot end temperature $T_h$ for the four substrates measured in this study.}
    \label{fig:QT-substrat}
\end{figure}

\begin{figure}
    \centering
    \includegraphics[width=0.95\linewidth]{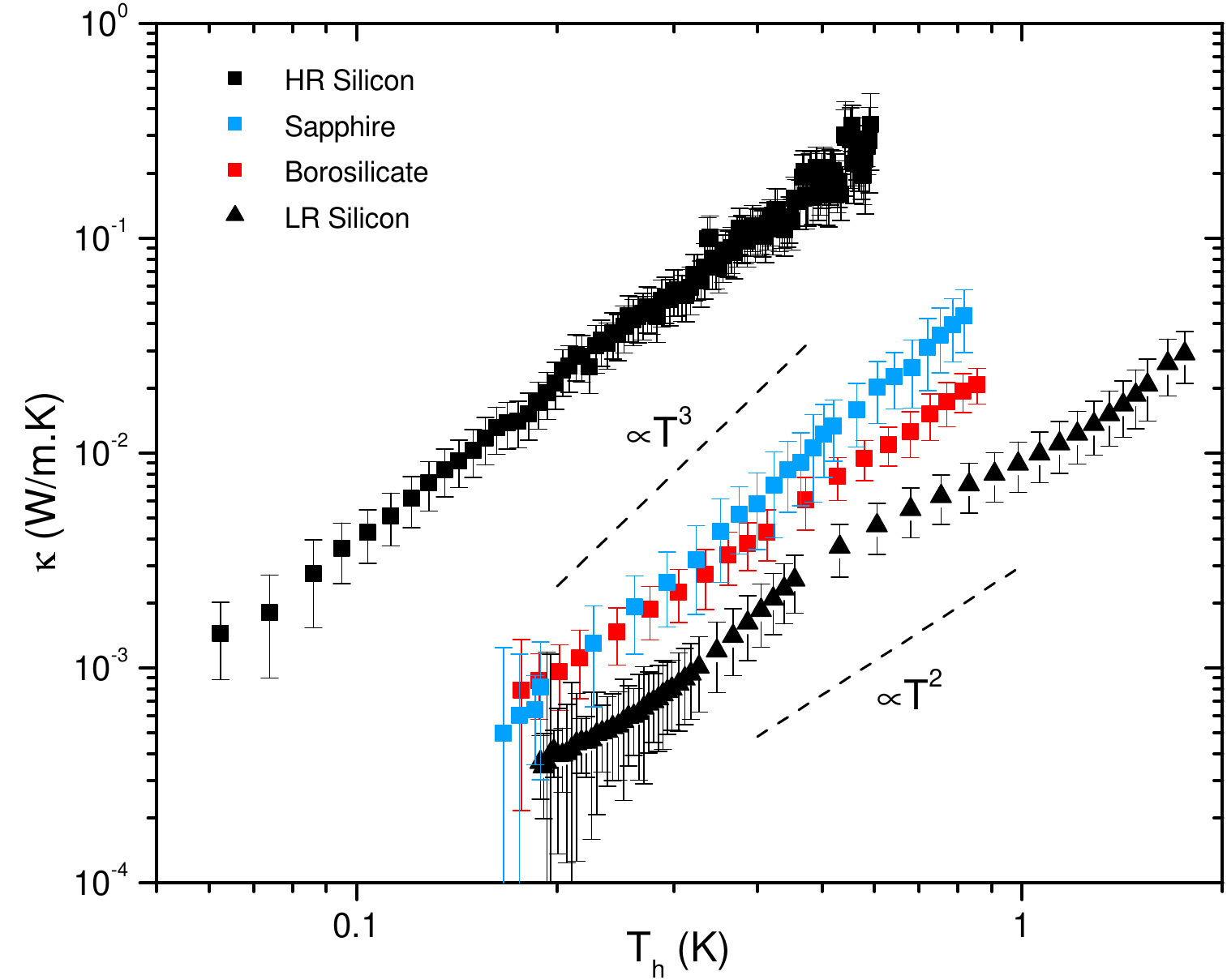}
    \caption{Effective thermal conductivity for all the measured substrates. Dashed lines are guide for the eyes.}
    \label{fig:kT-all}
\end{figure}

The measurement protocol is the following: we fixed the cryostat at a base temperature $T_0$ and applied a heating power $Q$ by steps of a few nanowatts with the heater. 
$T_0$ was typically set at 50~mK or 300~mK. For each heating step, we extract the equivalent rise in temperature by measuring $T_h$, the hot end temperature of the sample. By repeating this protocol for several $Q$, a large $Q(T_h,T_0)$ dataset is obtained and is presented in Fig.~\ref{fig:QT-substrat} for the four substrates.
The applied power ranges from a few nanowatts up to $\sim$10~µW, which is directly relevant to the context of cryo-CMOS integration: this upper bound is comparable to the power dissipated by control electronics, typically at several microwatts per qubit \cite{Cryo-CMOS_WATT}.
A description of the uncertainties estimation is given in the supplemental material \cite{SM}.
From these datasets, one can retrieve the effective thermal conductivity $\kappa$ using the steady state heat conduction definition:
\begin{equation}
     Q = \frac{wd}{L} \int_{T_0}^{T_h} \kappa(T) \,dT - Q_0
\end{equation}
where $Q_0$ is the parasitic heat load, assumed to be constant over the measured temperature range. As $T_0$ is kept constant during a measurement run, the effective thermal conductivity is obtained by differentiating \cite{TCI_method} $Q$ with $T_h$:
\begin{equation}
   \kappa(T_h) = \frac{L}{wd} \dfrac{dQ}{dT_h} 
\end{equation}
The differentiation is made using a numerical routine. The derivative at a given point is computed by taking the average of the slopes between the point and its two closest neighbours. The resulting thermal conductivities are presented in Fig.~\ref{fig:kT-all} and are discussed hereafter.

%----------------------------------------------------------------------------------------
%	Section
%----------------------------------------------------------------------------------------

\subsection{Substrate thermal conductivities}
%----------------------------------------------------------------------------------------
%	SILICON
%---------------------------------------------------------------------------------------
\subsubsection{Silicon}

We begin with the obtained thermal conductivities of the two silicon substrates.
A clear distinction is observed between the two substrate types. At 300~mK, the {HR} silicon exhibits a thermal conductivity of $5\cdot10^{-2}$~W/m$\cdot$K and a $T^{2.4}$ dependence with temperature. 
In principle, theory predicts a $T^3$ dependence at sufficiently low temperatures, where phonon transport is only limited by boundary scattering \cite{Casimir}. This deviation from the ideal $T^3$ law remains unexplained, nevertheless, it appears consistent with earlier literature reports where the $T^3$ scaling was also not retrieved exactly \cite{ThermalCondOfEl1972, Carruthers1957, White&Woods1956}.

For the {LR} silicon, its effective  thermal conductivity reaches only $8\cdot10^{-4}$~W/m$\cdot$K at 300~mK and the slope is close to $T^{2.2}$.
This two-order-of-magnitude reduction compared to {HR} silicon crystals was also observed in prior studies where doped silicon shows reduced thermal conductivities depending on the doping concentration \cite{Ashegi2002, Slack1964}.
%($\sim$0.1~W/m$\cdot$K at 2~K for heavily Boron-doped silicon \cite{Ashegi2002})
This reduction clearly highlights the influence of impurities on phonon-mediated heat transport. In the LR sample, boron atoms are expected to act as scattering centers for the phonon propagation, whereas in the HR sample, phonons propagate more freely, thus increasing the effective thermal conductivity.
In addition to the boron atoms, the Czochralski growth method used for the {LR} silicon wafer fabrication is known to introduce high oxygen impurity concentration (up to $7.3 \cdot 10^{17}$~cm$^{-3}$ in our case \cite{Si_Cz_Wafer}). Prior studies report that oxygen concentrations exceeding $10^{15}$~cm$^{-3}$ already lead to measurable reductions in phonon conductivity \cite{Si-Oxygen}, confirming the strong influence of oxygen-related defects.
In Ashegi~\textit{et~al.} \cite{Ashegi2002}, the low thermal conductivities measured for heavily doped samples were interpreted as a combination of oxygen contamination during the growth process and doping impurity concentrations.

Our experimental data follow the same general trends as those reported in the literature above 1~K for both pure \cite{ThermalCondOfEl1972, Carruthers1957, White&Woods1956} and doped silicon \cite{Ashegi2002, Slack1964}. For doped samples, however, the magnitude of the reduction depends strongly on doping concentration, and the available data are considerably scarcer than for bulk silicon.

%ADD T^2 and T^3

%----------------------------------------------------------------------------------------
%	SAPPHIRE
%----------------------------------------------------------------------------------------
\subsubsection{Sapphire}

For the sapphire, we measured a thermal conductivity of 2$\cdot10^{-3}$~W/m$\cdot$K at 300~mK, with a temperature dependence that closely follows a $T^3$ law.
While this $T^3$ scaling is consistent with sub-kelvin literature data \cite{Berman1955,WOLFMEYER1971,sahling1981}, the absolute magnitude of the conductivity in our sample is lower by nearly two orders of magnitude compared to the bulk values of approximately $10^{-1}$~W/m$\cdot$K at 300~mK \cite{Berman1955,WOLFMEYER1971,sahling1981}. 
This reduction may be attributed to size effects, given that the 536~µm-thick sapphire is the thinnest among the tested samples in this study, phonon propagation is strongly confined to the boundary scattering regime.
In our case, however, the larger suppression cannot be explained solely by the reduced substrate thickness.
A plausible explanation may lie in the surface state of the sample, as surface roughness and defects can further enhance diffuse boundary scattering \cite{Casimir,Holland}.
However, since the sample originates from a polished wafer, one of its faces is expected to exhibit negligible roughness compared to the dominant phonon wavelengths at sub-kelvin temperatures \cite{BOURGEOIS_MFP}. The pronounced reduction in thermal conductivity observed in our sapphire sample therefore remains to be understood. The mean free path analysis presented in Sec.~\ref{sec:mfp} provides additional physical  insight into this behavior.

%----------------------------------------------------------------------------------------
%	BOROSOLICATE
%----------------------------------------------------------------------------------------
\subsubsection{Borosilicate}

Finally, the borosilicate sample shows an effective thermal conductivity of approximately 2$\cdot10^{-3}$~W/m$\cdot$K at 300~mK, and following a $T^2$ dependence in the measured temperature range between 200~mK and 1~K, distinct from the $T^3$ behavior observed in the crystalline substrates studied above.
Our result closely reproduces prior studies on comparable glass like amorphous SiO$_2$ \cite{RAYCHAUDHURI1982, Raychaudhuri1989} and another borosilicate glass called BK7 \cite{Rosenberg2000}, all of which display a similar $T^2$ scaling below 2~K.
This behavior is characteristic of amorphous solids, where phonon transport differs fundamentally from that in single crystals: boundary scattering no longer influences thermal conduction, and the transport in this temperature regime can be described with the tunneling two-level system model~\cite{TLS}.

\subsection{Mean free path analysis}\label{sec:mfp}
\begin{figure*}
    \centering
    \begin{overpic}[width =0.475\linewidth]{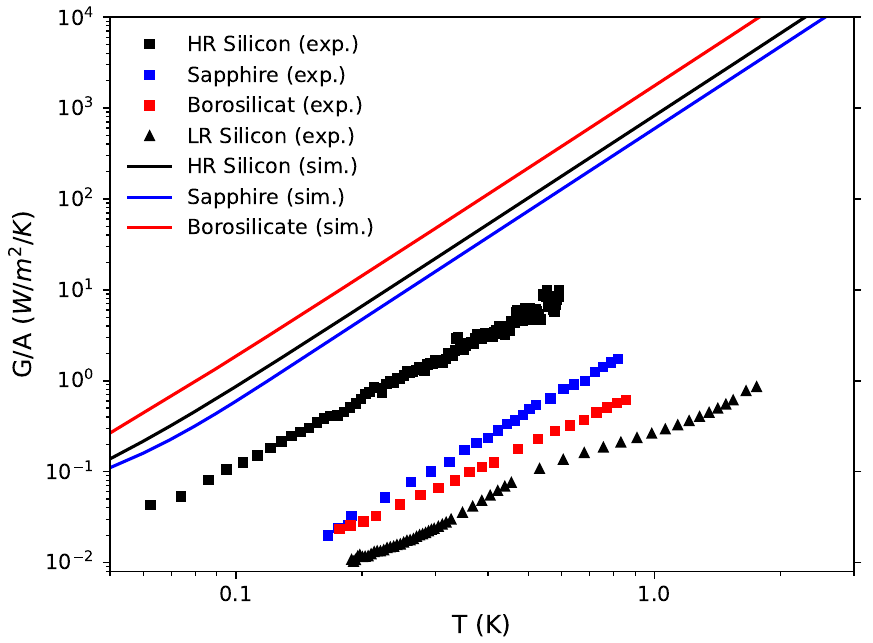}
        %\put(85.5,62){(a)} %Mettre (a) a droite
        \put(93,68){(a)}
    \end{overpic}
    \begin{overpic}[width =0.475\linewidth]{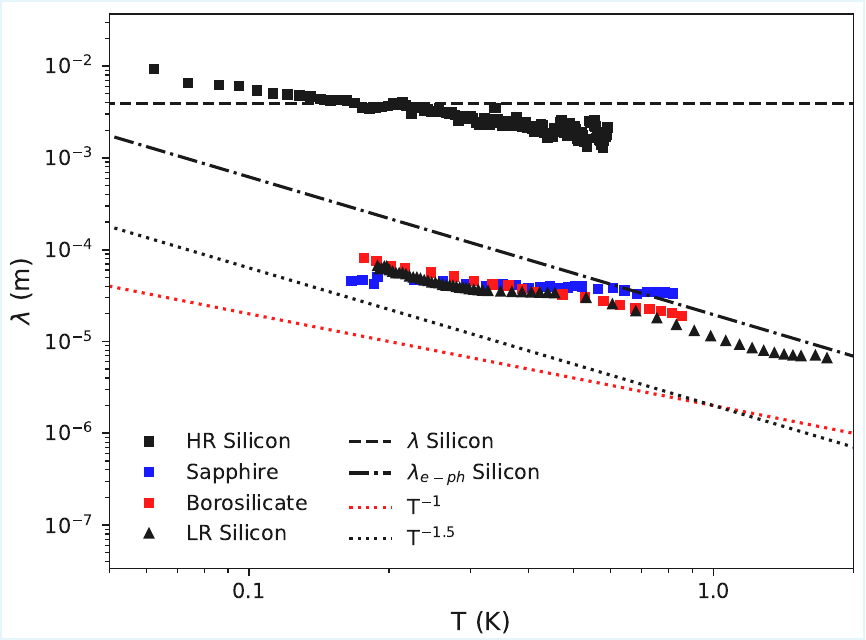}
        %\put(85.5,62){(b)}
        \put(93,68){(b)}
    \end{overpic}
    \caption{(a) Comparison of simulated ballistic conductance per area (continuous lines), $G_0/A$, with experimental measured conductance per area, $G/A$. (b) By using Eq.~\ref{mfp}, the mean free paths are extracted as function of temperature for various substrates. Horizontal dashed line is the Casimir boundary-limited mean free path obtained with Eq.~\ref{eq:lambda_geo}. The dash-dotted black line is the mean free path of phonon limited by hole-phonon scattering for the LR silicon substrate. Dotted lines are guides for T$^{-1}$ (red) and T$^{-1.5}$ (black) behavior.}
    \label{fig:sim}
\end{figure*}

The thermal conductivities reported above reflect the combined influence of material purity and sample geometry, but do not directly reveal the underlying phonon scattering mechanisms. To further interpret these results, we extract the phonon mean free path (MFP) from our experimental data and ballistic conductance simulations.

The ballistic thermal conductance is independent of the sample length $L$ and scales linearly with the cross-sectional area $A$.
In this work, the ballistic conductance per unit area, $G_{0}/A$ [Fig.~\ref{fig:sim}(a)], is evaluated using a finite-element non-equilibrium Green's function (FENEGF) approach~\cite{Polanco_2023}, based on the elastic constants and mass densities listed in Table~\ref{tab:elastic_cst}.

\begin{table}[h]
\caption{\label{tab:elastic_cst}Mass density $\rho$ (kg/m$^3$) and elastic constants (GPa) for Si, SiO$_2$, and sapphire. For sapphire, the values in parentheses correspond to the constants along the principal axis.}
\begin{ruledtabular}
\begin{tabular}{lccccc}
Material & $\rho$ & $c_{11}$ ($c_{33}$) & $c_{12}$ ($c_{13}$) & $c_{44}$ & $c_{14}$ \\
\hline
Si                      & 2330 & 165.6         & 63.9          & 79.5  & --   \\
SiO$_2$                 & 2200 & 77.5          & 15.7          & 30.9  & --   \\
Sapphire~\cite{Tefft1966} & 3986 & 500.1 (502.4) & 161.7 (111.4) & 151.0 & -23.3 \\
\end{tabular}
\end{ruledtabular}
\end{table}

When phonon transport is characterized by a finite mean free path $\lambda$, the corresponding conductance can be approximated by an interpolation between the ballistic and diffusive limits as~\cite{Datta_2005}
\begin{equation}
    G = G_0 \frac{\lambda}{L+\lambda},
    \label{mfp}
\end{equation}
where $G_0$ denotes the ballistic thermal conductance. This expression correctly recovers the ballistic limit $G \to G_0$ for $L \ll \lambda$ and the diffusive scaling $G \propto 1/L$ for $L \gg \lambda$.

Using the experimental conductance $G$ together with the calculated ballistic thermal conductance $G_0$, we extract the mean free path of phonons in the four substrates as a function of temperature from Eq.~\eqref{mfp}. The resulting temperature-dependent mean free paths are shown in Fig.~\ref{fig:sim}(b).

%------------------------------------
\subsubsection{HR Silicon}
For the high-resistivity silicon substrate, the extracted phonon mean free path is of the order of $4000$~µm, almost independent of temperature, in good agreement with the Casimir-type boundary-limited mean free path \cite{Casimir,BOURGEOIS_MFP},
\begin{equation}
    \lambda = 1.12 \sqrt{dw} = 3948~\text{µm},
    \label{eq:lambda_geo}
\end{equation}
where $w$ and $d$ denote, respectively, the width and thickness of the substrate, using the dimensions reported in Table~\ref{table:substrats}. This agreement indicates that phonon scattering is predominantly limited by diffuse boundary scattering at the substrate surfaces.

%------------------------------------
\subsubsection{LR Silicon}
For the low-resistivity silicon substrate, the extracted phonon mean free path is significantly shorter (30~µm at 500~mK) and exhibits an approximate $T^{-3/2}$ dependence. Such a temperature dependence is consistent with phonon-hole scattering, which is expected with the p-type boron doping.

For non-degenerate holes in silicon, assuming a parabolic valence band near the band edge, the acoustic phonon scattering rate can be estimated within the deformation-potential approximation \cite{Lundstrom_2000} as:
\begin{equation}
\frac{1}{\tau_{\mathrm{ac}}(E)}
\approx
\frac{D_{\mathrm{ac}}^{2} k_{B}T}{\pi \rho v_{s}^{2}\hbar^{4}}
\left(2m^{*}\right)^{3/2}\sqrt{E},
\end{equation}
where $D_{\mathrm{ac}}$ is the acoustic deformation potential, $\rho$ is the mass density, $v_{s}$ is the sound velocity, $m^{*}$ is the hole effective mass, and $E$ is the hole kinetic energy measured from the valence-band maximum. In the non-degenerate limit, the hole population follows Maxwell-Boltzmann statistics, so that a characteristic carrier energy is given by the thermal average,
\begin{equation}
\langle E \rangle = \frac{3}{2}k_{B}T.
\end{equation}
Using this estimate, the scattering rate scales as
\begin{equation}
\frac{1}{\tau_{\mathrm{ac}}} \propto T\sqrt{E}
\propto T^{3/2},
\end{equation}
after substituting $E \sim k_{B}T$ in the equipartition regime. Consequently, since the phonon mean free path is given by
\begin{equation}
\lambda = v_{s}\tau_{\mathrm{ac}},
\end{equation}
one obtains the characteristic scaling:
\begin{equation}
\lambda \propto T^{-3/2}.
\end{equation}

By substituting $D_{\mathrm{ac}} = 2.2$~eV \cite{Jacoboni_1977}, $v_{s} = (2v_{t}+v_{l})/3 = \mathrm{6700}$~m/s for the averaged sound velocity, $m^{*} = 0.81\,m_{0}$, and $\rho = 2330$~kg/m$^{3}$, the estimated mean free path ($\sim$50~µm at 500~mK) is found to be in good agreement with the value extracted from experiment, as shown in Fig.~\ref{fig:sim}(b).

%------------------------------------
\subsubsection{Sapphire}
For sapphire, the extracted phonon mean free path is approximately $50~\mu\mathrm{m}$ and remains essentially independent of temperature over the measured range. This value is, however, significantly smaller than the Casimir limit, $\lambda~\approx~2840$~µm. Such a temperature-independent but strongly reduced mean free path suggests that phonon transport is limited by temperature-insensitive structural disorder, most likely grain-boundary scattering.
This interpretation is supported by the recent work of Bai \textit{et al.} \cite{2025_Sapphire_domain_size}, who measured crystallite sizes between 34~nm and 283~nm in a 25~µm-thick A-plane sapphire wafer using X-ray diffraction, and attributed the reduced thermal conductivity of their sample to the grain-boundary scattering. 
Although our sample differs in thickness and crystal orientation, the presence of small crystallites in our C-plane wafer would be consistent with the reduced mean free path observed here.

%------------------------------------
\subsubsection{Borosilicate}
For borosilicate glass (approximated as pure SiO$_2$ for the simulation, despite its actual composition of 81\% of SiO$_2$, 13\% of B$_2$O$_3$, 4\% of Na$_2$O/K$_2$O and 2\% of Al$_2$O$_3$ \cite{Borofloat_wafer}), the extracted mean free path decreases with increasing temperature and remains far below the Casimir limit over the whole temperature range. This behavior indicates that phonon transport is not in the ballistic regime. It is also important to note that borosilicate is an amorphous material. In such a case, the applicability of a continuum phonon model, which implicitly assumes an underlying crystalline medium, is less clear. Although the ballistic conductance is estimated here from the elastic constants, this description may not fully capture the microscopic nature of thermal transport in an amorphous solid.

\subsection{Comparative discussion}
To summarize, among the substrates, the high-resistivity silicon is the most thermally conductive ($5\cdot10^{-2}$~W/m$\cdot$K at 300~mK), consistent with the boundary-limited phonon transport identified in the MFP analysis. 
Sapphire and borosilicate glass show similar values at the same temperature (2$\cdot10^{-3}$~W/m$\cdot$K at 300~mK), though their distinct temperature dependences ($T^3$ for crystalline sapphire versus $T^2$ for amorphous borosilicate) are expected to lead to diverging behavior at higher temperatures. In sapphire, the MFP analysis further suggests that structural disorder limits phonon transport.
The low resistivity silicon substrate shows the lowest conductivity ($8\cdot10^{-4}$~W/m$\cdot$K at 300~mK), reflecting the strong phonon-hole scattering as identified by the $T^{-3/2}$ MFP dependence. 
These results illustrate how substrate choice can be tailored for the application's thermal needs. 
In SOC architectures, for example, a {LR} silicon substrate with its lower thermal conductivity may help thermally isolating qubits from their control electronics, assuming that the dissipated power remains sufficiently low.
On the other hand, {HR} silicon can offer better heat removal, advantageous for the efficient thermalization of a power dissipative control die. 
Overall, these complementary characteristics suggest that no single substrate optimally fulfills all requirements for cryogenic integrated systems. Instead, the most suitable choice will depend on whether thermal isolation, heat dissipation, or {RF} performance \cite{CEA_RF_substrats} constitutes the primary constraint.

%----------------------------------------------------------------------------------------
%	SECTION 
%----------------------------------------------------------------------------------------
\section{Impact of on-chip routing on the thermal conductance}
Building on the data acquired from the substrate thermal conductivity study, we now investigate the impact of on-chip routing on the thermal conductance of samples made on {LR} silicon substrates, as this is the most used and promising platform for quantum technologies \cite{Quobly_monolithic_2025,ReviewSpinQu2023}. 
As discussed in the introduction, the use of superconducting routing is particularly advantageous in this context, as it suppresses the electronic contribution to heat transport, unlike normal metal routing which would add a significant parallel thermal path.
To this end, a planar test vehicle incorporating superconducting niobium (Nb) routing lines and gold (Au)-terminated pads is characterized, mimicking a simple SOC system.

\subsection{Test vehicle and experimental setup description}
\begin{figure}
    \centering
    \begin{overpic}[width=0.95\linewidth]{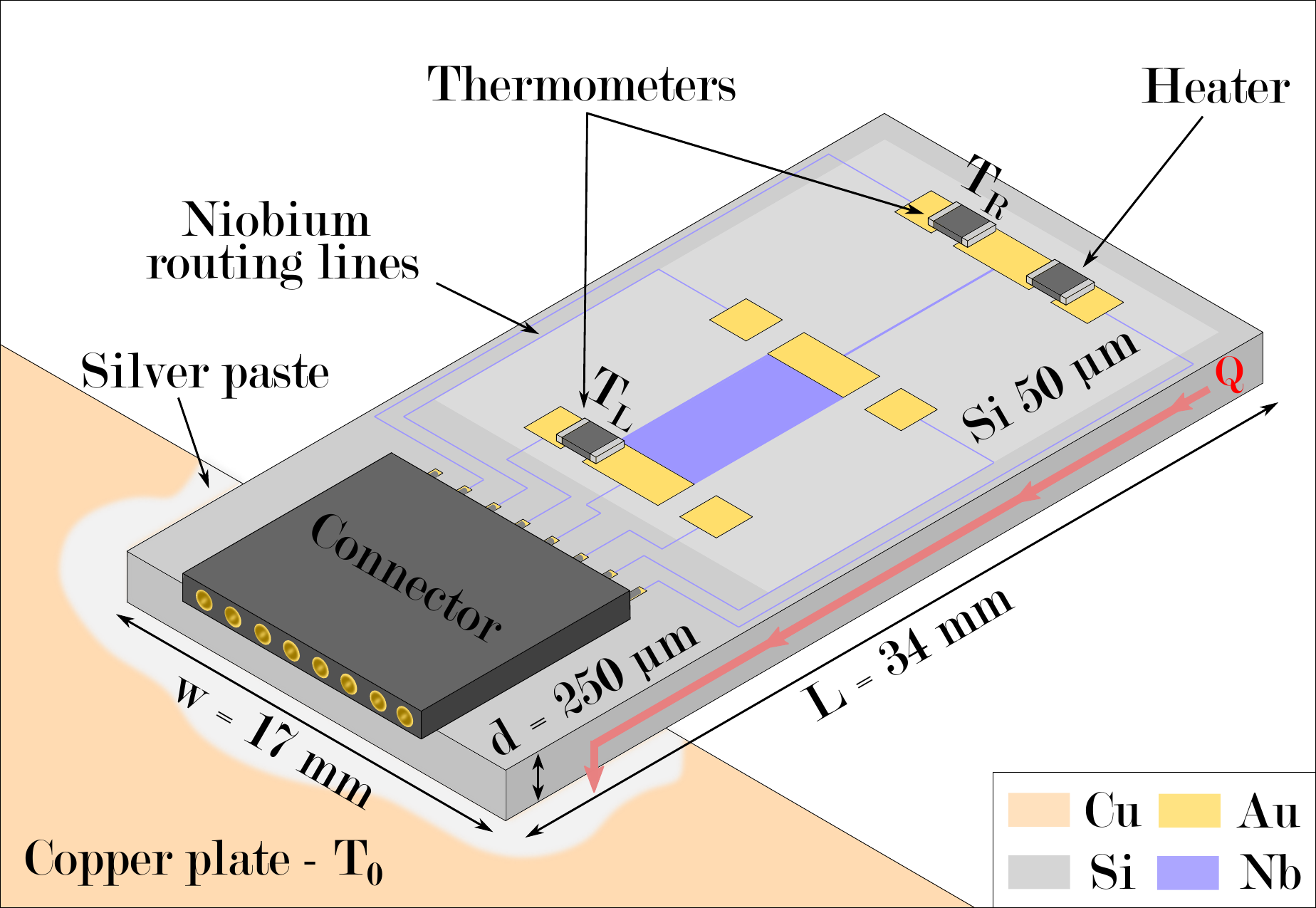}
        \put(1,65){(a)} % adjust coordinates if needed
    \end{overpic}
    
        % ---- Second figure with (b) ----
    \begin{circuitikz}[line width=1pt]
    \ctikzset{bipoles/thickness=1.2}
     % Label (b) top-left
    \node at (-8,1) {\small (b)}; % adjust coordinates as needed
    % Input node T0
    \draw (-8,0) node[circle, fill=black, inner sep=1.5pt] {} 
          node[above=0.1cm] {$T_{0}$};

    % Resistor chain with more spacing (reversed order)
    \draw (-8,0) to[R, l={$R_{b\,Si/Cu}$}] (-6,0)
            to[R, l={$R_{Si}$}] (-4.5,0)
          to[R, l={$R_{Routing}$}] (-0.5,0)
          to[short, i<^={Heat $Q$}] (0.2,0); % Arrow reversed

    % Parallel resistor for Substrate short
    \draw (-4,0) -- (-4,-1) to[R, l={$R_{Si}$}] (-1,-1) -- (-1,0);

    % Vertical T_L line
    \draw (-4,0) -- (-4,0.6) node[circle, fill=black, inner sep=1.5pt] {} 
          node[above=0.6cm] at (-4,0) {$T_{L}$};

    % Vertical T_R line
    \draw (-1,0) -- (-1,0.6) node[circle, fill=black, inner sep=1.5pt] {} 
          node[above=0.6cm] at (-1,0) {$T_{R}$};

    \end{circuitikz}

    \caption[Experimental setup and equivalent thermal model for planar routing test vehicles.]{(a) Experimental setup for the thermal conductance study of an on-chip routing test vehicle. (b) Simplified thermal equivalent model.} 
    \label{fig:HOT1_th_scheme}
\end{figure}
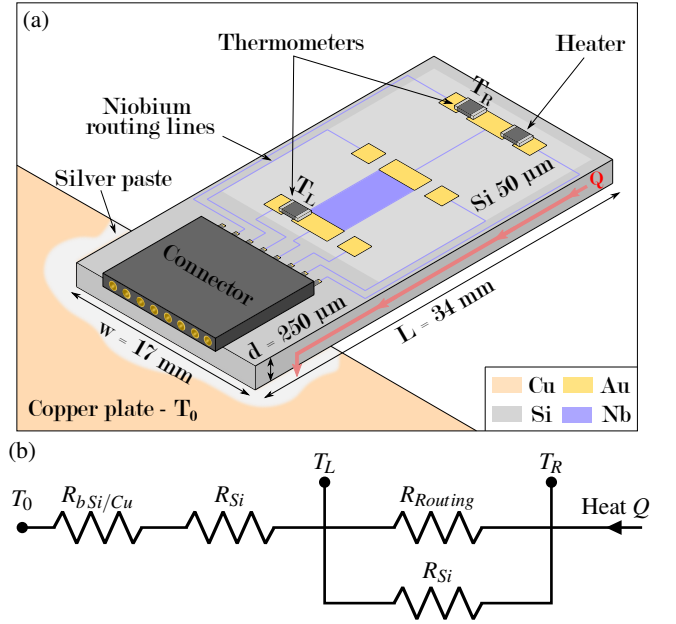

The on-chip routing test vehicle is illustrated in Fig.~\ref{fig:HOT1_th_scheme}(a).
It consists of three Au pads for thermalization purposes, which are connected by Nb routing lines. The superconducting routing and Au pads were fabricated following the processes presented in Ref.~\onlinecite{Candice_quantum}. 
The pads are deliberately elongated to promote thermal equilibrium, which is essential for treating the transport as one-dimensional.
In addition to the main pads, smaller Au pads are patterned at their vicinity to host the heater and the thermometers.
RuO$_2$ thermometers are soldered on the Au pads, enabling measurement of the local temperature at the left ($T_L$) and right ($T_R$) sections of the test vehicle.
A resistive heater is placed on the rightmost pad, to generate a controlled thermal gradient.
Although the design permits placing the heater also on the left and middle pads for flexibility, in practice, only the right-sided heater setup was used to establish a longitudinal thermal gradient. 
At the far left of the chip, a surface-mounted connector provides the necessary electrical connections. 
The link between the left and middle pads consists of a continuous Nb plane, while the middle and right pads are bridged by a single 4~µm wide routing line. Note that the Nb routing is encapsulated in a SiO$_2$ passivation layer, which is not represented in Fig.~\ref{fig:HOT1_th_scheme}(a) for clarity.

To accurately extract the thermal conductance of the routing, it is imperative to diminish the parallel thermal conductance of the underlying Si substrate. We achieved this by fabricating a backside cavity beneath the routing and Au pads region.
The wafer thickness was first reduced to 250~µm, after which the cavity was patterned by photolithography and etching techniques to a residual thickness of 50~µm. 
This reduction in the substrate's cross-section limits its thermal flow contribution, which should help maximizing the thermal gradient magnitude between $T_R$ and $T_L$. Based on the thermal conductivity of the LR silicon previously characterized, this residual Si volume should be sufficiently small to permit the establishment of a measurable thermal gradient and thus the measure of the sub-kelvin thermal conductance of the system.
More information about the cavity fabrication and packaging procedures are detailed in the supplemental material \cite{SM}.

The experimental setup is similar to the one described in Sec.~\ref{Sec:Substrats}, with the thermal equivalent model depicted in Fig.~\ref{fig:HOT1_th_scheme}(b). The considered thermal resistances include the Si/Cu thermal boundary resistance ($R_{b\,Si/Cu}$), the substrate resistance ($R_{Si}$), and the parallel routing resistance ($R_{Routing}$).
For simplicity, this model neglects the contribution of the thin ($\sim$500~nm) SiO$_2$ passivation layer present on top of the silicon substrate, whose thermal resistance is absorbed into the the $R_{Routing}$ term.
Furthermore, the thermal resistance arising from electron-phonon decoupling \cite{Hot_e_metals_1994} in the Au pads is also excluded. While present at low temperatures, we estimated that this decoupling effect is only significant below 100~mK (see observability criterion in Ref.~\onlinecite{Hot_e_metals_1994}), which remains below the temperature range investigated here (300~mK to 1.2~K).
From this equivalent model, the in-plane thermal conductance $G$ between $T_L$ and $T_R$ can be expressed as $G=1/R_{Routing} + 1/R_{Si}$.

\subsection{Thermal conductance measurement}

\begin{figure*}
    \centering
    \begin{overpic}[clip, trim={0 11.7cm 0 0}, width=0.475\linewidth]{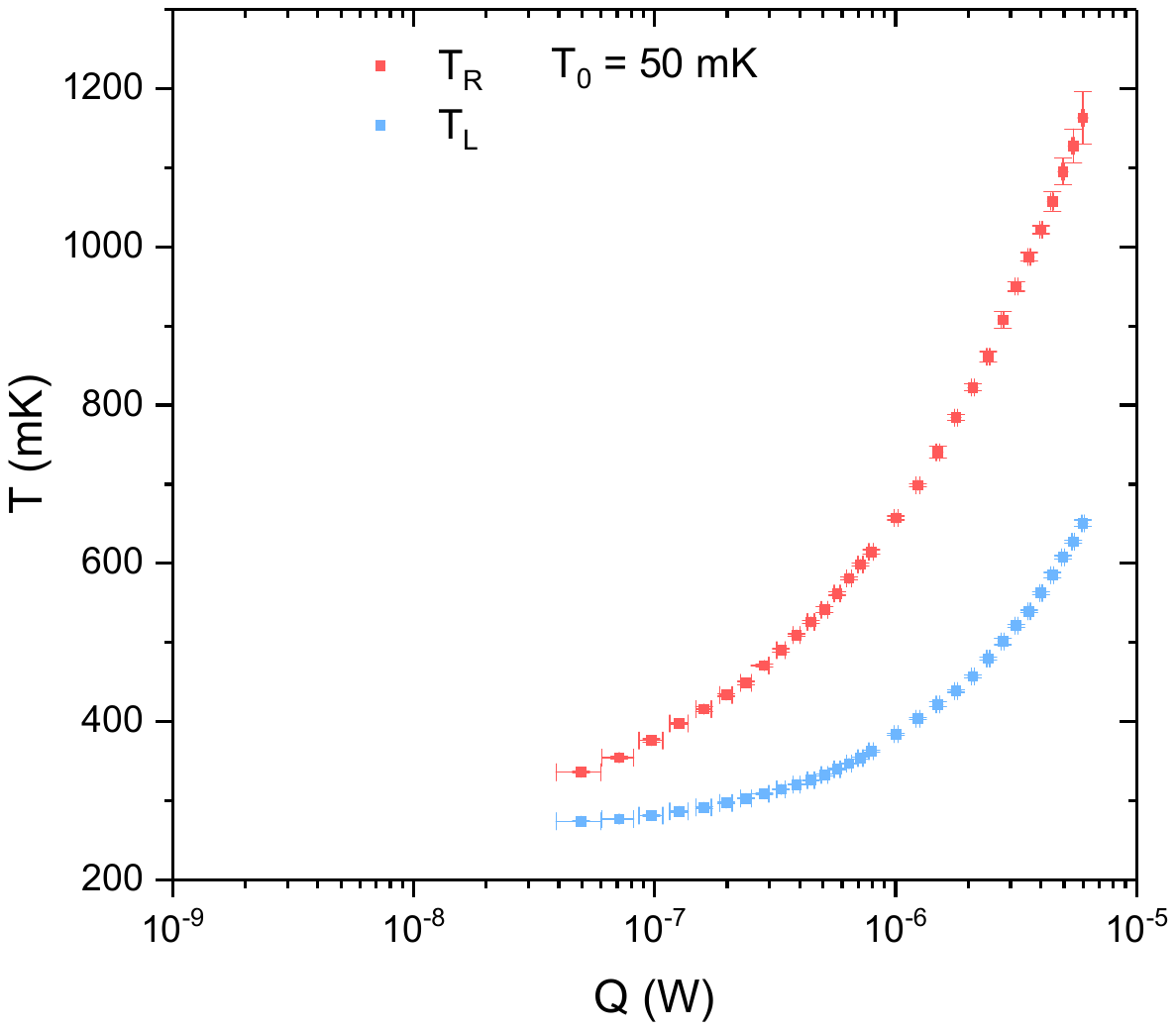}
        \put(21,77){(a)}
    \end{overpic}
    \begin{overpic}[clip, trim={0 11.7cm 0 0}, width=0.475\linewidth]{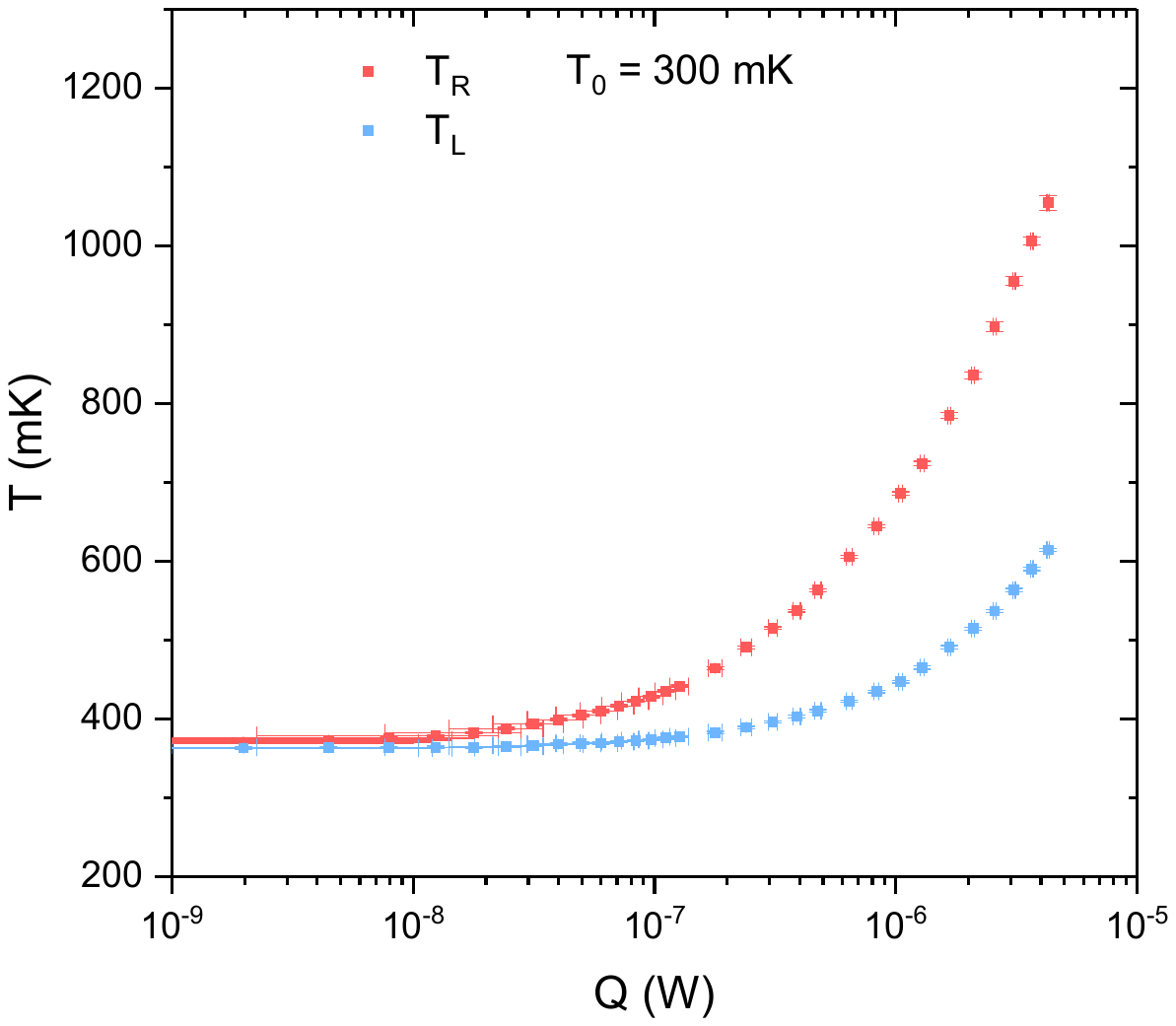}
    \put(21,77){(b)}
    \end{overpic}
    \caption{$T_L$ and $T_R$ thermometer temperatures as a function of the applied power $Q$ for two cryostat base temperatures: (a) at $T_0=~50$~mK and (b) at $T_0=~300$~mK. For both measurement runs, a thermal gradient is established between $T_R$ and $T_L$.}
    \label{fig:BT_Nb_T(Q)}
\end{figure*}

Two measurement runs were made, the first at the constant cryostat temperature $T_0=50$~mK and the second at $T_0=300$~mK. Both are presented in Fig.~\ref{fig:BT_Nb_T(Q)} which exhibits the $T_R$ and $T_L$ temperatures as a function of the applied heating power $Q$.
In both runs, a parasitic heat input $Q_0$ is present when the voltage source is connected to the heater. This unintended heating is believed to originate from a parasitic voltage drop across the heater resistor, potentially induced by the shared electrical ground configuration.
At $T_0=50$~mK [Fig.~\ref{fig:BT_Nb_T(Q)}(a)], $Q_0$ raised the sample temperature to approximately 300~mK, limiting the useful measurement range at this base temperature.
At $T_0=300$~mK [Fig.~\ref{fig:BT_Nb_T(Q)}(b)], the relative impact of $Q_0$ is reduced, raising the sample temperature only to 360~mK.
In both cases, this $Q_0$ remains a limiting factor for precise thermal characterization in this setup. Nevertheless, a thermal gradient is established above $\sim10$~nW of applied power in both measurement runs.

%\begin{figure}[h]
%    \centering
%    \includegraphics[clip, trim={0 11.7cm 0 0}, width=0.9\linewidth]{T(Q)_sample3_Nb_50mK.pdf}

%    \includegraphics[clip, trim={0 11.7cm 0 0}, width=0.9\linewidth]{T(Q)_sample3_Nb_300mK.pdf}
%    \caption{Measurement result on the thermal characterization of the on-chip routing test vehicle (a) at $T_0=~50$~mK and (b) $T_0=~300$~mK. For both measurement runs, a thermal gradient is obtained between $T_R$ and $T_L$.}
%    \label{fig:BT_Nb_T(Q)}
%\end{figure}

The obtained temperature difference $\Delta T =T_R-T_L$ is plotted as a function of the applied heat power in Fig.~\ref{fig:BT_Nb_G(T)}(a). 
From this dataset, the thermal conductance $G$ between $T_R$ and $T_L$ is computed using the linear approximation method, $G=Q/\Delta T$ 
and is evaluated at the mean temperature $(T_L+T_R)/2$. Using this method, the thermal conductance obtained for both measurement runs is plotted in Fig.~\ref{fig:BT_Nb_G(T)}(b). Both runs are showing consistent conductance results, consolidating the reproducibility of the measurement. 
At 500~mK the measured thermal conductance is $3\cdot10^{-6}$~W/K and exhibits a $T^{2.2}$ power-law temperature dependence.
Note that this scaling behavior closely matches the $T^{2.2}$ dependence observed for the LR silicon in the previous section, over the same temperature range.

\subsection{Discussion}
\begin{figure*}
    \centering
    \begin{overpic}[clip, trim={1cm 0.5cm 1cm 1cm}, width=0.475\linewidth]{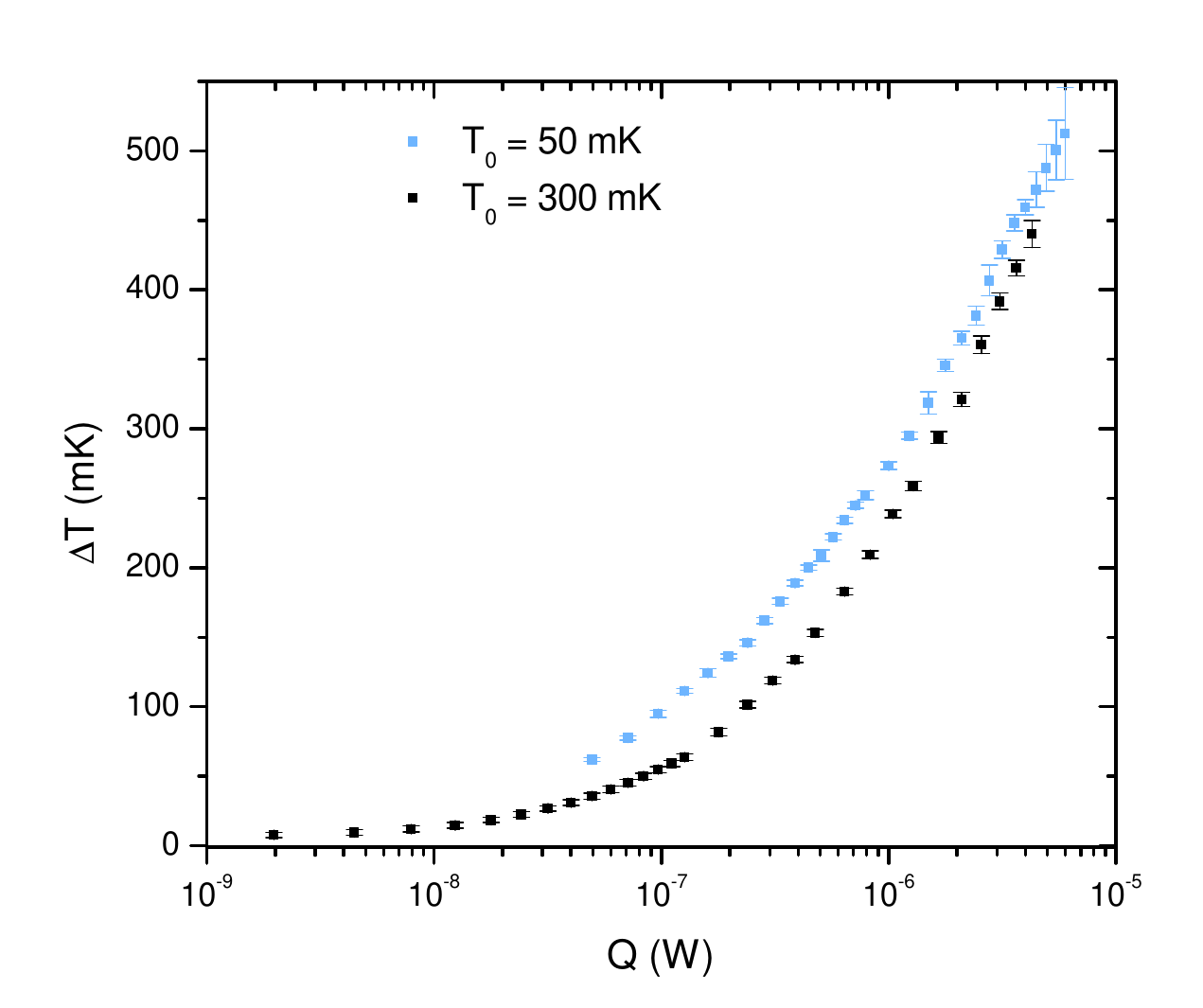}
        %\put(85.5,62){(a)} %Mettre (a) a droite
        \put(17,77){(a)}
    \end{overpic}
    \begin{overpic}[clip, trim={1cm 0.5cm 1cm 1cm}, width=0.475\linewidth]{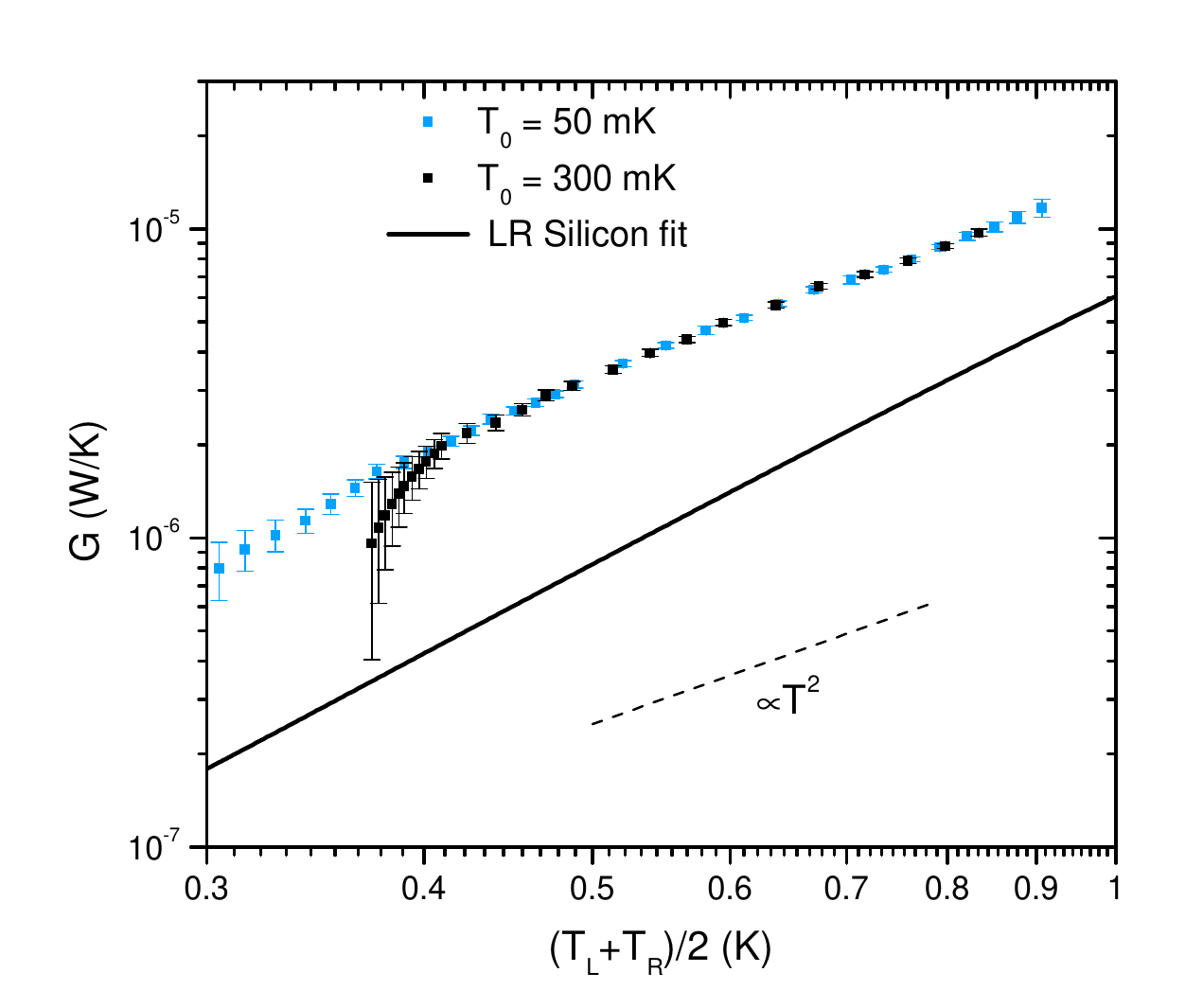}
        %\put(85.5,62){(a)} %Mettre (a) a droite
    \put(17,77){(b)}
    \end{overpic}
    \caption[Thermal conductance result of the Nb-routing test vehicle \#3.]{(a) Obtained temperature difference between the right $T_R$ and left $T_L$ thermometer versus the applied heat power $Q$. (b) Calculated thermal conductance $G=Q/\Delta T$ as a function of the mean temperature $(T_R+T_L)/2$ for the two measurement runs. The continuous line is the expected conduction from the LR silicon substrate only. The thickness used for the calculation was 250~µm, corresponding to the edge of the device, outside of the backside cavity. Dashed line is a $T^2$ guide for the eyes.}
    \label{fig:BT_Nb_G(T)}
\end{figure*}

To interpret the measured thermal conductance of the on-chip routing test vehicle, we compare it to the expected conductance of an identical device made solely of LR silicon. This is done via the relation $G_{Si,LR}=\kappa_{Si,LR} \, wd/L$ using the thermal conductivity determined in Sec.~\ref{Sec:Substrats}.
This calculation is added in Fig.~\ref{fig:BT_Nb_G(T)}(b) in the form of a continuous line using the device dimensions and assuming an effective thickness $d$ of 250~µm to account for the unetched silicon framing of the backside cavity.
The comparison demonstrates that the on-chip routing test vehicle exhibits a higher thermal conductance than if it was purely made out of {LR} silicon. 
For instance, at 500~mK, the expected conductance of the {LR} silicon is approximately $8\cdot10^{-7}$~W/K which is nearly four times lower than our  measurement.
We attribute this enhancement to the presence of the Nb routing lines and the Au-pads, which act as additional parallel thermal paths. 
Isolating the individual contributions of the routing and the pads will require systematic measurements (out of the scope of this paper) across varying device geometries, such as systematically modifying the routing line density and width, and comparing the resulting conductance.

A further critical observation from our measurements is that sample overheating became detectable at very low applied powers: a measurable temperature gradient is established between 1 and 10~nW at 300~mK for the substrate samples (Fig.~\ref{fig:QT-substrat}), and close to 10~nW for the on-chip test vehicle (Fig.~\ref{fig:BT_Nb_T(Q)}). 
Given that the samples are of centimeter-scale dimensions ($\sim$3$\times$2$\times$0.07~cm$^3$), this result highlights the extremely low power density required to perturb the thermal equilibrium. 
This provides an order-of-magnitude estimate of the dissipation tolerable by co-integrated cryo-CMOS before affecting nearby qubits in the same die.
It should be noted, however, that in our setup the samples are suspended, which does not accurately reproduce real device integration \cite{HorseRidge_CMOS, CMOS_wired_qubit_2025, CryoCMOS_Pauka2021}. In realistic architectures, the die is glued over its full surface to the cryostat to maximize thermal exchange, which would likely increase this dissipation threshold. Ultimately, the challenge of thermally shielding qubits from co-integrated dissipative electronics can be understood as a balance between in-plane thermal conduction (through the substrate and routing lines) and vertical heat removal toward the cryostat, which is predominantly restricted by the thermal boundary resistance \cite{SwartzPohl} at the chip-cryostat interface.
The literature provides additional context to this last observation at sub-kelvin temperatures.
Savin~\textit{et al.}~\cite{Pekola_Thermal_Budget} estimated that a 5$\times$5~mm² silicon chip can only accommodate both active electronics and sensitive elements (qubits) if power dissipation remains below $\sim$50~nW.
Another recent study from Blagg~\textit{et~al.} \cite{Overheating_Si_2022} reported that the local temperatures on the silicon chip, microns away from the heat source, are found to increase by the order of 100~mK with only 100~nW of applied power.
Keep in mind that those results should be considered as order-of-magnitude estimates, subject to variations arising from device geometry, material choice, and thermal contact.
Despite this, one thing is certain: with current levels of power dissipation in cryo-CMOS, typically several microwatts per qubit~\cite{Cryo-CMOS_WATT}, protecting qubits from this generated heat when both are co-integrated within the same substrate, even separated by a few mm, represents a significant thermal management challenge. 
Even assuming state-of-the-art demonstrations, projecting less than 100~nW dissipation for a million-transistor chip \cite{2025_low_power}, the margin between this dissipation and the acceptable heat load on qubits is virtually non-existent. 
This thermal management bottleneck strongly motivates the co-integration of quantum and classical circuits within SIP architectures, where thermal decoupling can be enhanced through the use of 3D interconnect technologies \cite{VTT_superconducting_assemblies,REVIEW_3D_Integration_for_quantum}.
Alternatively, moving the dissipative electronics to a higher temperature stage of the dilution refrigerator \cite{2026_Si_Quantum_HRL,2025_Rev_Classical_interfaces}, combined with the use of thermally resistive superconducting flex \cite{DeLaBroise, 2026_Si_Quantum_HRL}, would constitute an appealing solution.

\section{Conclusion}

This work has addressed the thermal characterization of SOC-like planar structures relevant to cryogenic integrated systems. We first examined the thermal conductivities of four representative substrates, showing that substrate choice is an important parameter for the thermal management of integrated systems. High-resistivity silicon, for instance, provides enhanced thermal conductivity, while low-resistivity silicon exhibits a two-order-of-magnitude reduction at sub-kelvin temperatures, making it more favorable for thermal insulation purposes.
Mean free path extraction via FENEGF simulations further suggested the dominant scattering mechanisms in each substrate: boundary-limited transport in HR silicon, hole-phonon scattering in LR silicon, structural disorder in sapphire, and amorphous-regime transport in borosilicate.

The last study on the on-chip routing test vehicle gives some insights when combined with the substrate thermal conductivity study. 
Although the routing was found to increase the thermal conductance by a factor of four compared to the substrate of our test vehicle, this enhancement was only measurable because the silicon thickness had been locally reduced to 50 µm. In a standard SOC architecture, where the substrate retains its full thickness (typically several hundred micrometers), the substrate thermal conductance would largely exceed the routing contribution.
In a sense, the effective thermal conductance of planar cryogenic integrated circuits can, to a first approximation, be reduced to the thermal conductivity of its substrate alone.

The overall conclusion from this work is that in SOC architectures, where qubits and control electronics would share the same substrate, the dissipation tolerable before significantly perturbing the qubit temperature is of the order of tens of nanowatts at 300 mK for centimeter-scale dies, which is far below the microwatt-level dissipation of current cryo-CMOS \cite{Cryo-CMOS_WATT}. This observation encourages function partitioning through 3D integration and system-in-package approaches as a means to place a large number of cryo-CMOS circuits in close proximity to the qubits while mitigating thermal crosstalk. This motivates the development of advanced technologies such as highly thermally resistive interconnects \cite{JL_MC,CBM_SSL,Pablo_MC_bonding} and thermally resistive substrates \cite{2026BraggReflectorsPhonon}.

\begin{acknowledgments}
This work was supported (in part) by the French National program “Programme d’investissement d’avenir, IRT Nanoelec, n° ANR-10-AIRT-05’’.
The authors gratefully acknowledge Simon Zihlmann for kindly providing the sapphire samples and Olivier Bourgeois for valuable discussions.
The authors would like to thank the CEA Si platform and the layout team.
\end{acknowledgments}

%\bibliographystyle{apsrev4-2}
\bibliography{biblio.bib}

%SUPPLEMENTAL MATERIAL FOR ARXIV FORMAT

\clearpage
\onecolumngrid

\renewcommand{\thefigure}{S\arabic{figure}}
\renewcommand{\thetable}{S\arabic{table}}
\renewcommand{\theequation}{S\arabic{equation}}
\setcounter{figure}{0}
\setcounter{table}{0}
\setcounter{equation}{0}

\begin{center}
{\Large \textbf{Supplemental Material}}
\end{center}

\setcounter{section}{0}
\renewcommand{\thesection}{S\Roman{section}}

%------------------------
%NEW SECTION
%------------------------
\section{Measurement test on Silicon substrate for the validation of the experimental setup/analysis method}

\begin{figure}[h]
 % ---- First figure with (a) ----
    \centering
    \begin{overpic}[width=0.7\linewidth]{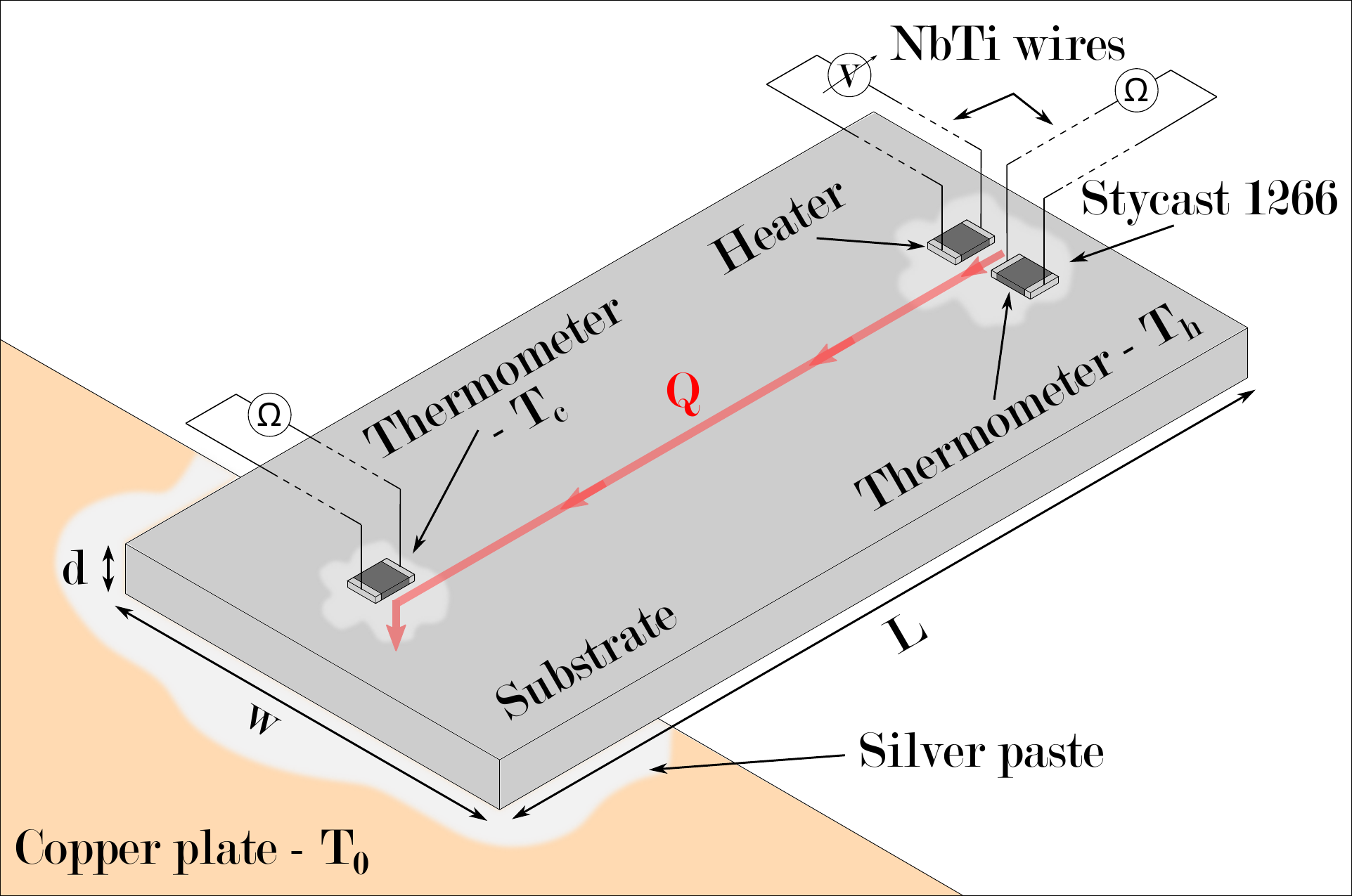}
        \put(6,61){(a)} % adjust coordinates if needed
    \end{overpic}
    % ---- Second figure with (b) ----
    \begin{circuitikz}[line width=1pt]
    \ctikzset{bipoles/thickness=1.2}
     % Label (b) top-left
    \node at (-8,1) {\small (b)}; % adjust coordinates as needed
    % Input node T0
    \draw (-8,0) node[circle, fill=black, inner sep=1.5pt] {} 
          node[above=0.1cm] {$T_{0}$};

    % Resistor chain with more spacing (reversed order)
    \draw (-8,0) to[R, l={$R_{b\,Substrate/Cu}$}] (-5,0)
          to[R, l={$R_{Substrate}$}] (-3,0)
          to[short, i<^={Heat $Q$}] (-1,0); % Arrow reversed

    % Vertical T_c line
    \draw (-5.2,0) -- (-5.2,0.6) node[circle, fill=black, inner sep=1.5pt] {}
          node[above=0.6cm] at (-5.2,0) {$T_{c}$};
    
    % Vertical T_h line
    \draw (-3,0) -- (-3,0.6) node[circle, fill=black, inner sep=1.5pt] {}
          node[above=0.6cm] at (-3,0) {$T_{h}$};

    \end{circuitikz}
    \caption{(a) Experimental setup for the thermal conductivity study of various substrates. (b) Simplified equivalent thermal model of the setup.}
    \label{figSM:Substrat_Sample}
\end{figure}

A validation measurement was performed on the high-resistivity (HR) silicon sample to verify that the experimental method effectively captures the thermal resistance of the substrate itself, and that the thermal bottleneck is not located at the substrate/copper plate interface due to the thermal boundary resistance.
To this end, we used the experimental setup shown in Fig.~\ref{figSM:Substrat_Sample}(a). The only difference with the main study setup is the presence of an additional thermometer: the cold-end thermometer $T_c$, located above the copper plate/substrate interface. 
The equivalent thermal model is modified accordingly and is described in Fig.~\ref{figSM:Substrat_Sample}(b). By probing $T_c$, we can evaluate $R_{Substrate}$, the thermal resistance arising from the substrate material itself, independently from  $R_{b\,Substrate/Cu}$: the thermal boundary resistance between the substrate and the copper plate.

\begin{figure}
    \centering
    \includegraphics[clip, trim=0.5cm 1cm 3.2cm 1cm, width=0.475\linewidth]{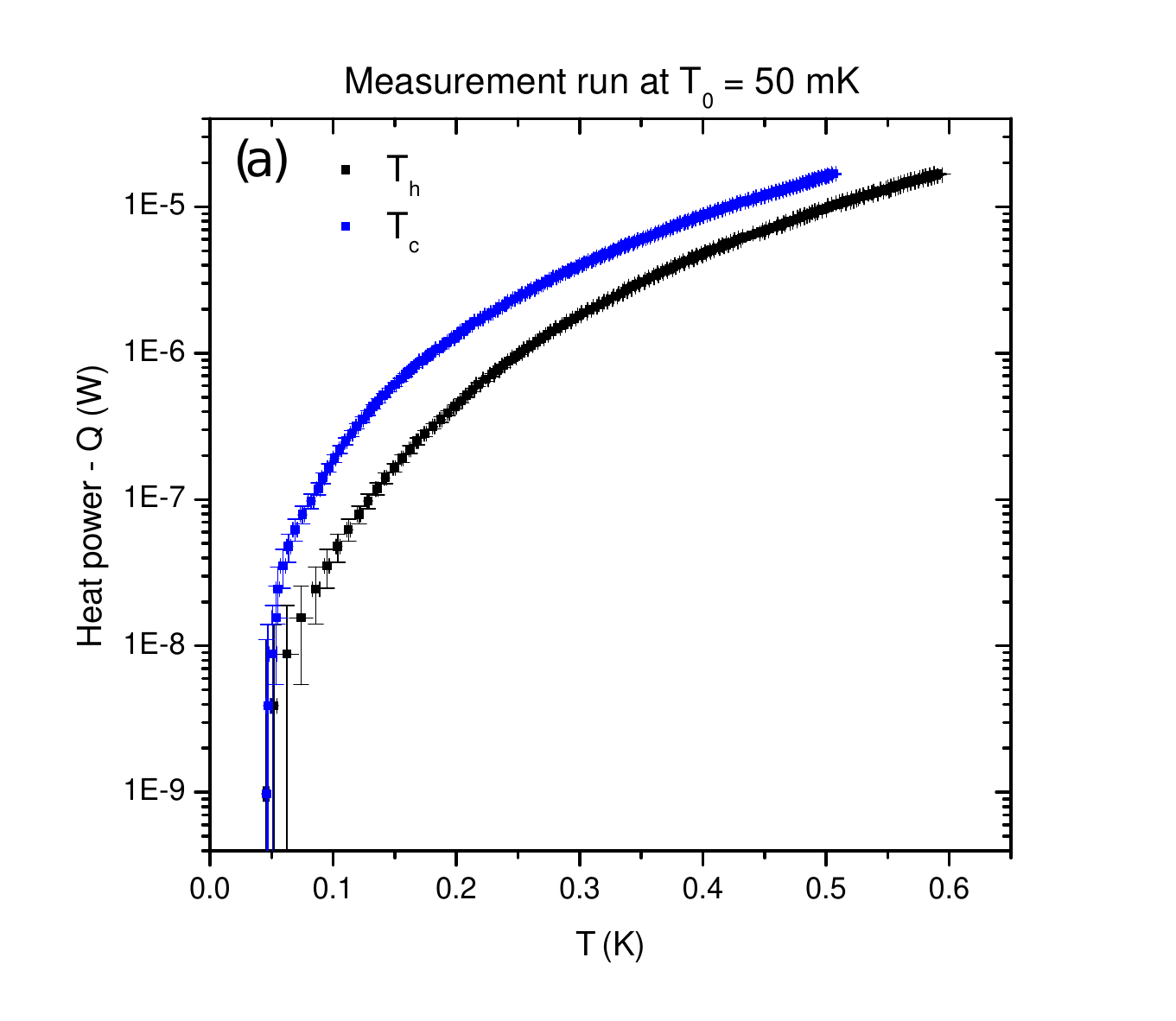}
    \hfill
    \includegraphics[clip, trim=0.5cm 1cm 3.2cm 1cm, width=0.475\linewidth]{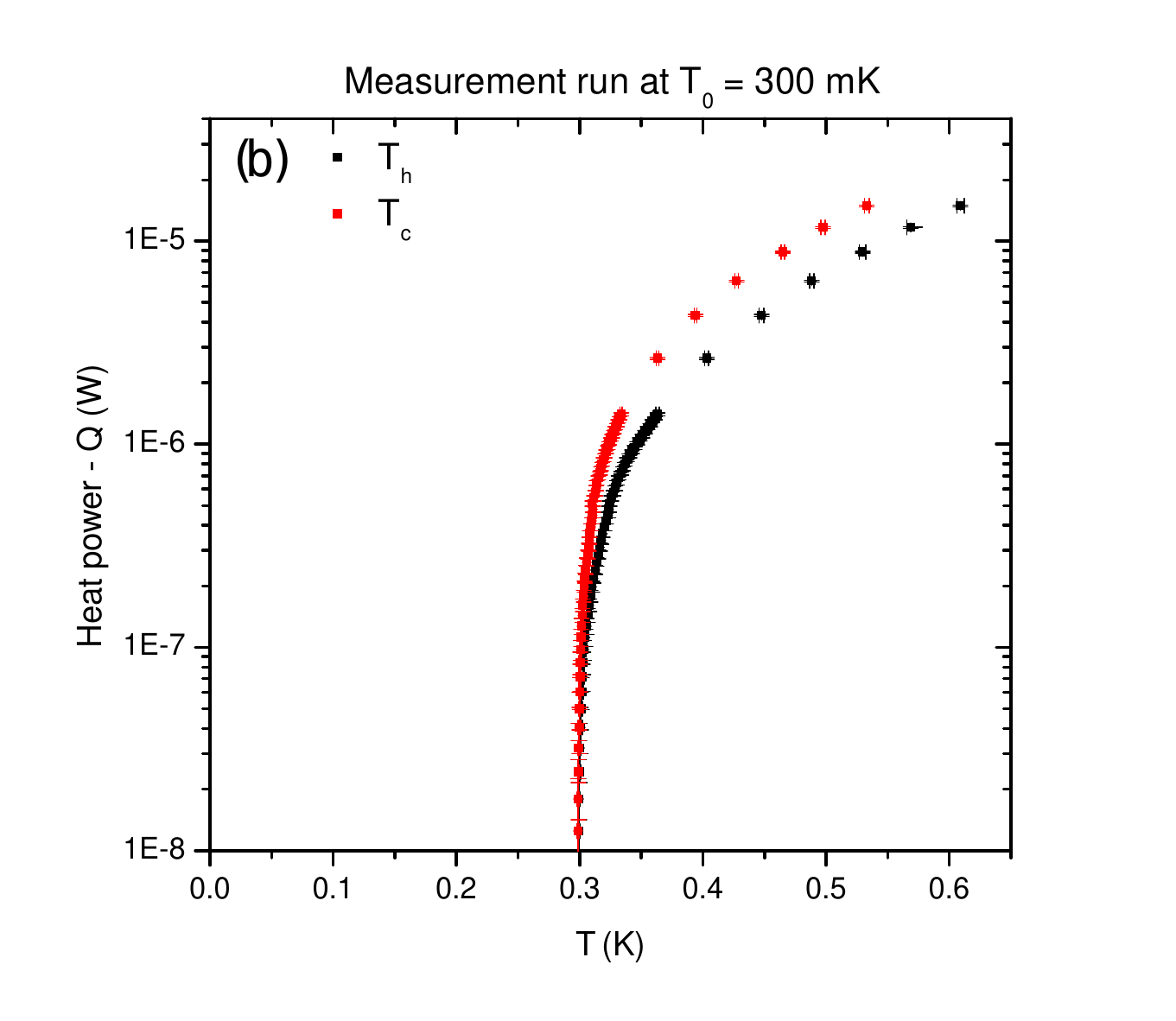}
    \caption{Applied heat power $Q$ versus temperatures ($T_h$ and $T_c$) for the two measurement runs performed at (a) $T_0=$~50~mK and (b) $T_0=$~300~mK.}
    \label{fig:SM_Q(T)}
\end{figure}

Using this setup, two measurement runs were performed at constant cryostat temperatures of $T_0=$~50~mK and $T_0=$~300~mK. The obtained heating power versus temperature for each thermometer is shown in Fig.~\ref{fig:SM_Q(T)}.
From these $Q(T_h,T_c,T_0)$ datasets, the thermal conductance $G$ can be extracted using two analysis methods.

The first is the linear approximation method applied between $T_c$ and $T_h$, assuming a temperature-independent thermal conductivity and a small temperature gradient:
\begin{equation}
    G(T_{avg})= \frac{Q}{T_h-T_c}
\end{equation}
where $T_{avg}=(T_h+T_c)/2$.

The second method is the thermal conductivity integral (TCI) method \cite{TCI_method}, as used in the main study, applied between $T_h$ and $T_0$.
Since $T_0$ is fixed, the conductance can be obtained by differentiating $Q$ with respect to $T_h$:
\begin{equation}
   G(T_h) = \dfrac{dQ}{dT_h} 
\end{equation}
The differentiation is made using a numerical routine. The derivative at a given point is computed by taking the average of the slopes between the point and its two closest neighbours.

\begin{figure}
    \centering
    \includegraphics[clip, trim=0.5cm 1cm 3.1cm 1cm, width=0.54\linewidth]{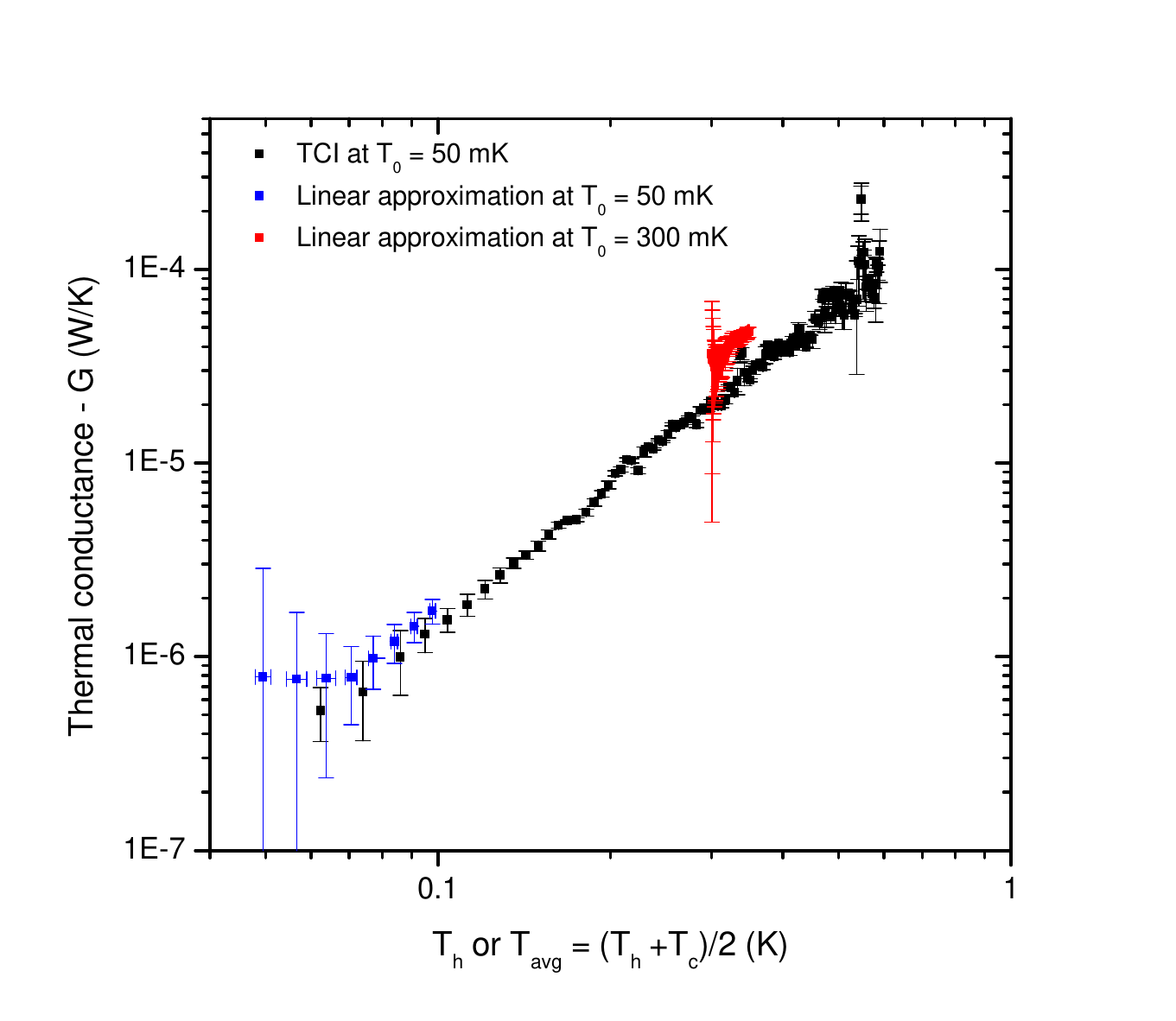}
    \caption{Comparison between the two methods for the extraction of the thermal conductance, either the {TCI} (using only $T_0$ and $T_h$) or the linear approximation method (using  $T_0$, $T_c$ and $T_h$).}
    \label{fig:SM_G(T)}
\end{figure}

The thermal conductances obtained from both methods are compared in Fig.~\ref{fig:SM_G(T)}. For clarity, only the measurement run at $T_0=$50~mK was used for the TCI method, while both runs at $T_0=$50~mK and $T_0=$300~mK were used for the linear approximation method. 
For the latter, only data points at low applied power (up to a few nanowatts) were retained, consistent with the assumption of a small temperature gradient and temperature-independent thermal conductivity.

Within the measurement uncertainties, both methods yield consistent conductance values. Since the TCI method includes the interfacial contribution $R_{b\,Substrate/Cu}$ while the linear approximation method does not, the agreement between the two results demonstrates that $R_{b\,Substrate/Cu}$ is negligible compared to $R_{Substrate}$. For the remaining samples in the study, it was therefore decided that a single hot-end thermometer combined with the TCI method is sufficient to accurately extract the thermal conductance of the substrate.
Additionally, the TCI method offers clear practical advantages: it extends the usable range of applied power $Q$ by removing restrictions on the temperature difference between $T_h$  and $T_c$, and it requires one fewer thermometer.

%------------------------
%NEW SECTION
%------------------------
\section{Thermal conductance uncertainties}\label{Apendix_Uncer}

\subsection{Sample thermometer calibration}
\begin{figure}[h]
    \centering
    \includegraphics[clip, trim=1.9cm 3cm 3cm 4cm, width=0.5\linewidth]{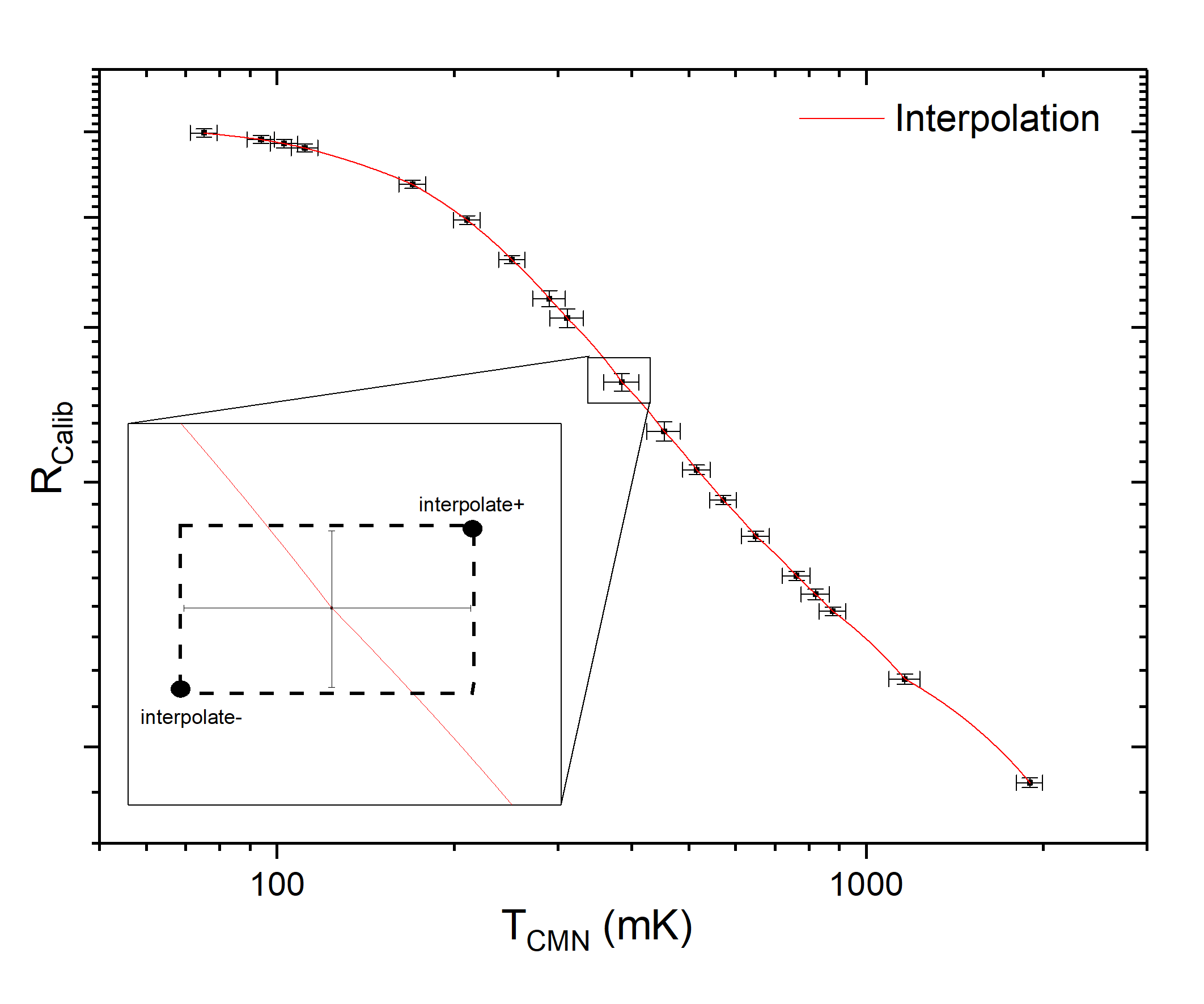}
    \caption[Typical sample's thermometer calibration dataset with interpolation.]{Typical sample's thermometer calibration dataset with interpolation. The uncertainty $\delta{T_{CMN}}$ is typically 5\%, while $\delta R_{calib}$ = 0.1\% + standard variation during the temperature step measurement. }
    \label{fig:ex.calib}
\end{figure}

The first source of uncertainty comes from the calibration of the sample thermometer. The latest is calibrated during the cryostat cool-down or heat-up by making several long temperature steps. The equivalent sample temperature measured during this temperature step is obtained by comparison with the main thermometer of the cryostat: a powdered Cerium Magnesium Nitrate (CMN) thermometer. The uncertainty coming from this principal CMN thermometer $\delta{T_{CMN}}$ is roughly estimated to 5\%. The measured electrical resistance $R_{calib}$ of the sample thermometer during the calibration has an uncertainty $\delta R_{calib}$ of (0.1\% + standard variation during the temperature step). 
With this, we obtain the typical calibration plot in Fig.~\ref{fig:ex.calib}. From this dataset and during the thermal conductivity measurement, the measured sample temperature $T$ is obtained from the measured $R$ by interpolating the latest on the calibrated dataset $R_{calib}(T_{CMN})$.

\subsection{Thermometer uncertainty during experiment $\delta R$ and $\delta T$}
During the experiment, the measured electrical resistance $R$ has an uncertainty coming from the measurement itself and the standard deviation $\sigma_R$ during the heating step (constant temperature),
\begin{equation}
    \delta R = 0.1\%\cdot R + \sigma_R
\end{equation}
Next, to obtain the uncertainty $\delta T$, two more interpolations are extracted from the calibration dataset, corresponding to the extrema of the calibration curve (see inset in Fig.~\ref{fig:ex.calib}). 
Thanks to this, the extrema of $\delta T$ can be calculate using:
\begin{align}
        & \delta T^- = |\text{interpolate}^-( R + \delta R) - T| \\
        & \delta T^+ = |\text{interpolate}^+( R - \delta R) - T|
\end{align}
Then $\delta T$ is simply the maximum value between $\delta T^-$ and $\delta T^+$.

\subsection{Applied heat power uncertainty $\delta Q$}
Joule heating is used for the input heat power in the experiment with a  200~$k \Omega$ resistor. The heater resistance $R_h$ is considered stable in the temperature range studied, so the only uncertainty associated is its nominal value. We have estimated that
\begin{equation}
    \delta R_h \sim 1\%\cdot R_h = 2000\Omega
\end{equation}
The heater is biased using a voltage source (SN8310), the associated uncertainty from the datasheet on the applied voltage $U$ is:
\begin{equation}
    \delta U =0.007\%\cdot U+20\mu V
\end{equation}
Finally, $\delta Q$ is estimated by propagating the uncertainties:
\begin{equation}
    \delta Q = Q_0 + \sqrt{\left(\dfrac{\partial Q}{\partial U}\right)^2(\delta U)^2 
                + \left(\dfrac{\partial Q}{\partial R_h}\right)^2(\delta R_h)^2}
\end{equation}
with $Q_0$ being the parasitic heat load coming from the environment. The latter parameter is estimated to be constant over the temperature range and was found to mainly come from the voltage source. $Q_0$ was estimated to be $\sim10~nW$ by making two calibrations with the same thermometer with and without the voltage source plugged into the heater. 

\subsection{Thermal conductance uncertainty $\delta G$}
Lastly, the uncertainty of the measured thermal conductance $\delta G$ can be obtained from all the previous uncertainties and by applying the uncertainty propagation formula, knowing that $G= \partial Q/\partial T$:
\begin{equation}
    \delta G = \sqrt{\left(\dfrac{\partial G}{\partial T}\right)^2(\delta T)^2 
                + \left(\dfrac{\partial G}{\partial Q}\right)^2(\delta Q)^2}
\end{equation}

%------------------------
%NEW SECTION
%------------------------
\section{Assembly and packaging of the on-chip routing test vehicle}\label{sec:Packaging}
\begin{figure}[h!]
\centering
\includegraphics[width=0.9\linewidth]{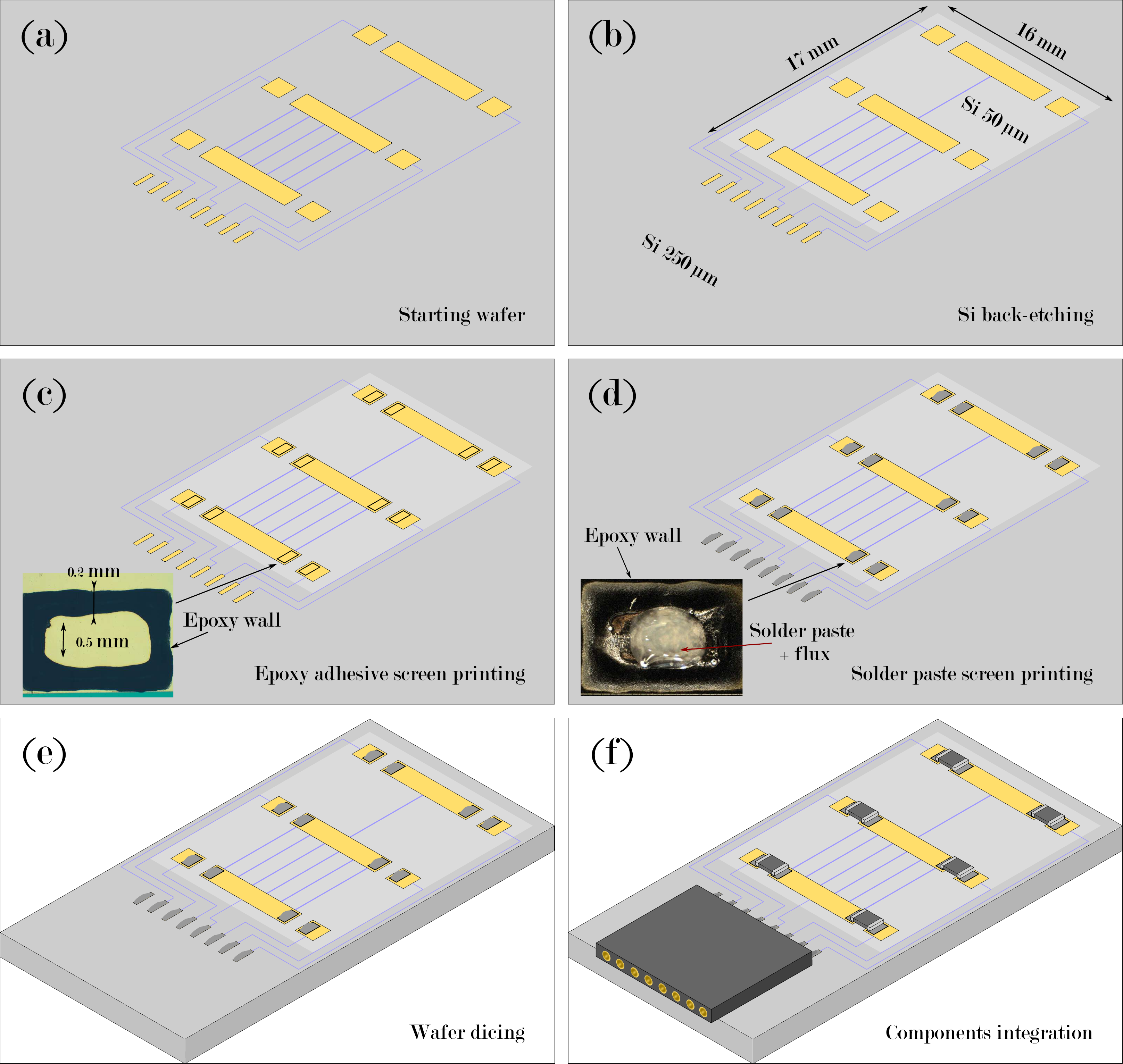}
\caption[Packaging workflow of the planar routing test vehicle.]{Packaging workflow of the planar routing test vehicle. Beginning with the processed wafer (a), the substrate is thinned down and locally back-etched (b). Next, two screen printings are made: the first for the epoxy adhesive (c), the second for the solder paste (d). Finally, the wafer is diced (e), and the connector and {SMD} resistors are mounted (f).}
\label{fig:process_assem_BT}
\end{figure}

After the clean-room fabrication of the 200~mm wafers following the process presented in \cite{Candice_quantum}, the wafers were diced and packaged into test vehicles suitable for cryogenic thermal characterization. 
The main requirements were compatibility with a cryogenic environment and the direct integration of heaters and thermometers to allow generating and measuring thermal gradients. An additional objective was to implement plug-and-play electrical connectivity, to facilitate sample mounting in the cryostat and increase measurement throughput. To reduce the impact of the substrate on the thermal measurement, a cavity was etched on the backside to locally minimize its thickness.
The complete multi-step packaging procedure is depicted in Fig.~\ref{fig:process_assem_BT} and is described in the following.
\begin{itemize}
    \item[(a)] \textbf{Starting wafer} -- The packaging flow starts with the fully processed wafer. In Fig.~\ref{fig:process_assem_BT}, the blue lines represent the superconducting (Nb) routing level and the gold pads are in yellow. Note that the Nb routing is encapsulated in a SiO$_2$ passivation layer, which is omitted from the top views for clarity.

    \item[(b)]\textbf{Silicon back-etching} -- The first step aims to create a "membrane-like" cavity from the substrate backside, to minimize its participation in the thermal transport within the area of the test structure. 
    For this purpose, the original wafer is firstly thinned from 725~µm down to 250~µm using grinding methods and a chemical-mechanical polishing process for the last micrometers. Next, a cavity is formed in the remaining silicon substrate using standard photolithography and deep reactive ion etch techniques down to 50~µm, in a large surface area under the golden pads [see Fig.~\ref{fig:process_assem_BT}~(b)]. 
    Given the large etched area ($17\times16$~mm²), it was decided to keep a residual thickness of 50~µm under the test structure to ensure the mechanical stability of the test vehicle. The reason behind this is the intrinsic compressive stress between Si and SiO$_2$ that is known to induce warpage and even breakage if the silicon thickness is further reduced. On top of that, since the test vehicle will be settled in a cryogenic environment, the membrane must withstand the mechanical constraints caused by the temperature range and vibrations from the cryostat.
    A schematic cross-section of the test vehicle at this stage is shown in Fig.~\ref{fig:Cross_section}, illustrating the SiO$_2$ passivation layer and the backside cavity.

    \begin{figure}
        \centering
        \includegraphics[width=0.8\linewidth]{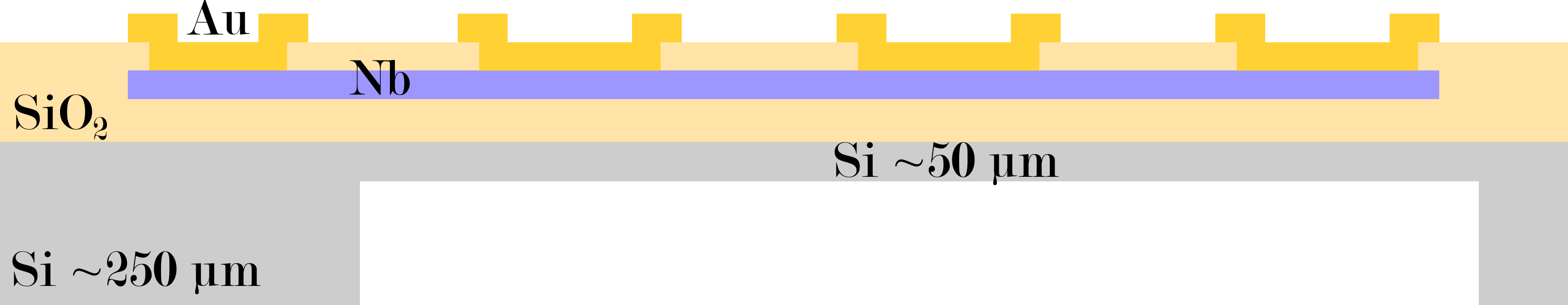}
        \caption{Schematic cross-section of the on-chip routing test vehicle after cleanroom fabrication and backside cavity etching (prior to the packaging steps described in Sec.~\ref{sec:Packaging}(c)–(f)). The SiO$_2$ passivation layer encapsulating the Nb routing is represented here but omitted from the top views in Fig.~\ref{fig:process_assem_BT} for clarity.}
        \label{fig:Cross_section}
    \end{figure}

    %The second part of the packaging process aims to solder the components needed for the measurements, which are the thermometers, the heaters and the electrical connector. To do that, solder paste was needed but one challenge was to develop a reproducible and repeatable process and to avoid the solder past to spread on the large pad area.... 

    \item[(c)]\textbf{Epoxy adhesive screen printing} -- The second part of the packaging process aims to solder the components needed for the measurements, namely thermometers, heaters, and electrical connectors. To do so, solder paste needs to be applied on the Au pads. However, as long Au pads are present in the design, there is a risk that the solder spreads all over these pads during the reflow, preventing it from properly contacting the mounted resistances. 
    To face this, the investigated solution consists of introducing solid epoxy structures in the form of hollow rectangles, whose function is to confine the solder paste and prevent it from spreading.
    To implement this solution, a wafer-level screen-printing approach was adopted, as it enables the processing of a larger number of samples with improved reproducibility while remaining compatible with 200~mm wafer technologies.
    However, the requirement for hollow rectangles introduces a limitation: conventional solid stencils, typically employed for screen printing, cannot be used. 
    This is because the central part of the hollow rectangles in the mask (which represents the negative of the epoxy result) needs to be self-supporting, which is impossible in a solid stencil.
    The solution found was to use a mesh stencil instead. Mesh stencils have a thin, grid-like structure that maintains mechanical support to the rectangular central part.
    Employing this technique, the epoxy is deposited on the wafer, followed by a 10-minute reflow at 160°C, ensuring proper epoxy curing. A result example is presented in the inset of Fig.~\ref{fig:process_assem_BT}(c). The fabricated epoxy walls exhibited a typical width of 0.2~mm, and an opening of approximately 0.5$\times$1~mm².

    %To implement this solution, a wafer-level screen printing of the epoxy is made using a mesh mask\footnote{All the screen-printing masks used in this work were fabricated by \href{https://www.koenen.de/}{Christian Koenen GmbH} with designs provided by our team at CEA.}. 
   % Because of the epoxy rectangles design, a mesh stencil is preferred compared to a classical 'solid' stencil: the center of the rectangle in the mask --- which represents the negative of the epoxy result --- needs to be self-supporting. This is only achievable with a thin, mesh-like structure that maintains mechanical support to the rectangle. Employing this technique, the epoxy is deposited on the wafer followed by a 10-minute reflow at 160°C ensuring proper epoxy curing. A result example is presented in Figure~\ref{fig:process_assem_BT}(c). 
    % \footnote{DELO DUALBOND OB793 epoxy, from \href{https://www.delo-adhesives.com/}{Delo-adhesives}}

    \item[(d)]\textbf{Solder paste screen printing} -- The next step is the solder paste deposition using another wafer-level screen-printing method. The solder paste is a Bi-Sn-Ag alloy, characterized by a low reflow temperature. Compared to a more common Sn-Ag-based solder paste, which requires a 240°C reflow, the Bi-Sn-Ag alloy allows us to limit the thermal budget and thus protects the previously deposited epoxy rectangles during its reflow. 
    The screen printing is realized using a 3D stainless steel stencil mask to accommodate the surface topology introduced by the epoxy rectangles. After the deposition, the soldering reflow is performed using a conveyor oven with a peak temperature of 180°C. The post-reflow result is presented in the inset of Fig.~\ref{fig:process_assem_BT}(d). 
    %\footnote{Indium5.7LT-1 Type 3, from \href{https://www.indium.com/}{Indium Corporation}}

    \item[(e)]\textbf{Wafer dicing} -- After these preparation steps, the wafers are diced into individual samples with a semi-automatic wafer dicing machine, with parameters adjusted to preserve the soldering and epoxy during the washing step.

    \item[(f)]\textbf{Components integration} -- The final step consists of mounting the required components for the thermal characterization. 
    Based on the design and targeted measurement configuration, seven of them are needed: three thermometers, three heaters, and one electrical connector. Both the thermometers and the heaters are surface-mounted devices ({SMD}) resistors, and two package sizes are compatible with the pad size; either 1206 or 0805. The connector is a single-row, surface-mounted, rectangular-shaped, 8-pin connector. %\footnote{171-008-8P Single Row Surface Mount Micro-D MicroStrips, from \href{https://www.glenair.com}{Glenair}}.
    Manual flux deposition is performed on the soldering using a dispensing tool before component placement. The latter is made using a pick-and-place machine or directly by hand, which is facilitated thanks to the stickiness of the flux. The sample with all the components is then annealed through a conveyor oven, using the same temperature range as for the solder reflow. 
    Optionally, to improve mechanical robustness, an epoxy can be further dispensed around the connector and is cured at 80°C for 30 minutes. 
\end{itemize}

\begin{figure}[h]
    \centering
    \includegraphics[width=0.9\linewidth]{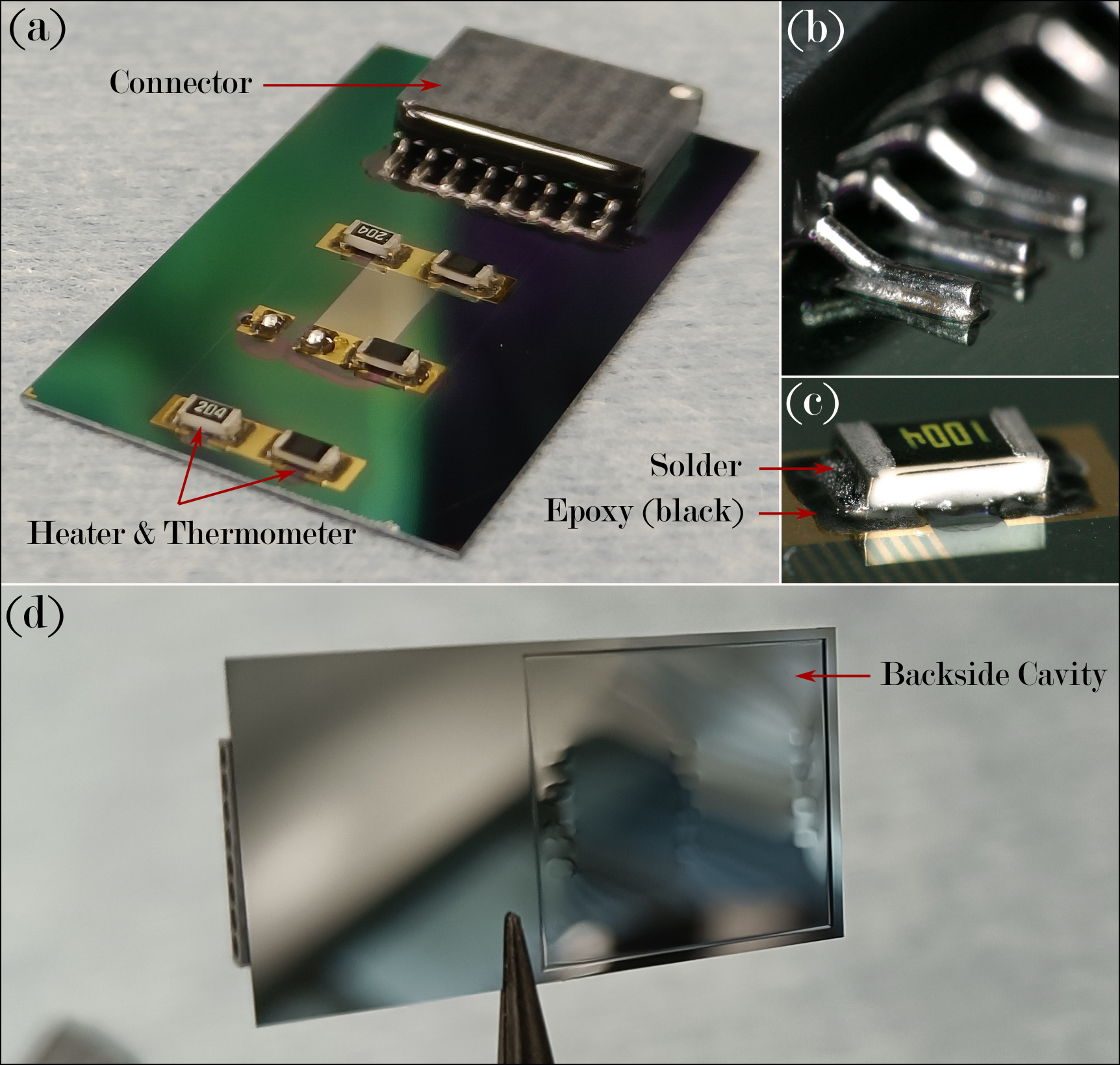}
    \caption[Pictures of the planar routing test vehicle.]{(a) Completed planar routing test vehicle. Zoom on (b) the 8-pin electrical connector and (c) one resistor with the black epoxy rectangle. (d) Sample backside, stress-induced buckling is observed on the 50~µm silicon membrane.}
    \label{fig:BT}
\end{figure}

The resulting test vehicle is pictured in Fig.~\ref{fig:BT}(a). One of the resistors, a heater in this context, is omitted as it was not essential for the thermal characterization, but its location was eventually added in the design for flexibility. Focusing on the resistor in Fig.~\ref{fig:BT}(c), one can perceive the epoxy rectangle in black that effectively hinders the solder paste from spreading during the reflow. 
On the substrate backside, shown in Fig.~\ref{fig:BT}(d), where the large etched rectangle defines the membrane. The created membrane exhibits distortion caused by mechanical stress.%, a phenomenon better known as \textit{buckling}. 
%In fact, for thin membrane fabrication purposes, this effect is typically mitigated through the use of counter-strategies based on compensating the stress to avoid the membrane's distortion \cite{Si_membranes}. Alternatively, low-stress silicon nitride is often favored to facilitate membrane fabrication \cite{SiN_membranes}. 
%While these methods remain relevant for future design improvements, we have already managed to build a compliant demonstrator that fulfills the immediate requirements: sensor integration, ease of handling, and plug-and-play usability.

%\bibliographystyle{apsrev4-2}
%\bibliography{biblio.bib}

\end{document}